Journal of Insurance and Financial Management, Vol. 1, Issue 4 (2016) 68-131

# JIFM
## JOURNAL OF INSURANCE AND FINANCIAL MANAGEMENT

# An Investigation of the Structural Characteristics of the Indian IT Sector and the Capital Goods Sector – An Application of the R Programming in Time Series Decomposition and Forecasting

Jaydip Sen [a,*], Tamal Datta Chaudhuri [b]

[a,b] Calcutta Business School, Bishnupur – 743503, West Bengal, India

ABSTRACT

Time series analysis and forecasting of stock market prices has been a very active area of research over the last two decades. Availability of extremely fast and parallel architecture of computing and sophisticated algorithms has made it possible to extract, store, process and analyze high volume stock market time series data very efficiently. In this paper, we have used time series data of the two sectors of the Indian economy – Information Technology (IT) and Capital Goods (CG) for the period January 2009 – April 2016 and have studied the relationships of these two time series with the time series of DJIA index, NIFTY index and the US Dollar to Indian Rupee exchange rate. We establish by graphical and statistical tests that while the IT sector of India has a strong association with DJIA index and the Dollar to Rupee exchange rate, the Indian CG sector exhibits a strong association with the NIFTY index. We contend that these observations corroborate our hypotheses that the Indian IT sector is strongly coupled with the world economy whereas the CG sector of India reflects India's internal economic growth. We also present several models of regression between the time series which exhibit strong association among them. The effectiveness of these models have been demonstrated by very low values of their forecasting errors.

*Corresponding Author:
jaydip.sen@acm.org

Journal of Insurance and Financial Management (ISSN: 2371-2112)

# 1. Introduction

The literature on portfolio choice and forecasting of stock returns has concentrated on various characteristics of companies like their Profitability Indicators, Leverage, P/E ratio, P/BV ratio, Size, Volume of Trade, Market Capitalization and Dividend Payout Ratio. Understanding companies by the above mentioned parameters, somehow, leads to some standardization and robs the companies of their individuality. Every company is distinct, and one of the sources of this distinctiveness stems from the sector to which it belongs. Each sector is tied to some aspects of the economy. These aspects may be socio economic and/or demographic characteristics, the income distribution pattern, the extent of global integration, domestic endowment of resources, and state of the technology or market size. The literature has not explicitly captured these aspects, and hence sectoral distinctions have not been modelled adequately. As a consequence, any methodology for forecasting of stock returns for a sample set of diverse companies, we feel, falls short of the desired results.

In our previous work, we have been emphasizing on this specific aspect of sectoral characteristics (Sen and Datta Chaudhuri, 2016a; Sen and Datta Chaudhuri, 2016b; Sen and Datta Chaudhuri, 2016c). Following our decomposition approach, we have demonstrated that indeed the sectors are different in terms of their trend, seasonal and random components. We have pointed out that, for India, each of the above components is tied to some social or economic feature, and the forecasting methodology that we suggested in our work incorporates these sectoral characteristics.

In this paper, we look at two different sectors in India, namely the Information Technology (IT) and the Capital Goods (CG) sectors, and demonstrate that these two sectors are completely different in terms of their behavioral characteristics. We hypothesize that while the IT sector, being a services sector, is tied to the rest of the world, the CG sector is very much tied to the India story. We use the R programming framework to decompose the time series of the IT and CG sectors stock market index into trend, seasonal and random components, and then relate the movement of each component to the components of Dow Jones Industrial Average (DJIA), NIFTY (Indian National Stock Exchange Index) and the US Dollar to Indian Rupee Exchange Rate. Our contention is that instead of comparing the movement of the aggregates, one should compare the movement of the components for better understanding of the sectors. This would give further insight into the choice of stocks for portfolio formation and also in portfolio redesigning.

The contribution of this work is threefold. First, we propose a time series decomposition approach and then illustrate that the proposed technique provides us with a deeper understanding of the behavior of a time series by observing the relative magnitudes of its three components namely trend, seasonal and random.

Second, we present mechanisms of studying associations among different time series using various graphical and statistical tests. The association analysis provides us further insights into the behavioral characteristics of different time series. Several hypotheses are also validated using the association analysis.

Third, we develop various models of regression for time series that exhibit strong associations among them in our study. The models are constructed using suitably designed training data sets and then tested using appropriate test data sets. The forecast accuracies of each of the models are computed so as to have an idea about their efficacies and robustness.

The rest of the paper is organized as follows. Section 2 briefly discusses the methodology in constructing various time series and decomposing the time series into its components. It also presents a brief outline on the forecasting frameworks designed in this work using the R programming language. Section 3 provides a detailed discussion on the methods of decomposition, the decomposition results of all the sectors under study, and an analysis of the results. Section 4 presents a methodology of comparing and analyzing association between several time series under our investigation. The association between the Indian IT sector and Indian CG sector with DJIA index, US Dollar to Indian Rupee exchange rate, and the NIFTY index are studied in great detail. We present several hypotheses and validate them through graphical means and several statistical tests. Section 5 presents several linear models for forecasting that enables one to forecast the index of one sector given the index of another sector to which is known to be strongly associated. We present eight models and present extensive results to demonstrate their efficacy and effectiveness in forecasting. In Section 6, we discuss some related work in the current literature. Finally, Section 7 concludes the paper.

## 2. Methodology

In this section, we provide a brief outline of the methodology that we have followed in our work. However, each of the following sections contains detailed discussion on the methodology followed in the work related to that Section. We have used the *R programming language* (Ihaka & Gentleman, 1996) for data management, data analysis and presentation of results. R is an open source language with very rich libraries that is ideally suited for data analysis work. We use daily data of the Indian IT sector index, Indian CG sector index, NIFTY index, DJIA index and the US Dollar to Indian Rupee exchange rate for the period January 2009 to April 2016. The daily index values are first stored in five plain text files – each sector data in one file. The daily data are then aggregated into monthly averages resulting in 88 values in the time series data. These 88 monthly average values for each sector are stored in five separate plain text files – each sector monthly average in one file. The records in the text file for each sector are read into an R variable using the *scan( )* function in R. The resultant R variable is converted into a monthly time series variable using the *ts( )* function defined in the *TTR* library in the *R* programming language. The monthly time series variable in R is now an aggregate of its three constituent components: (i) trend, (ii) seasonal, and (iii) random. We then decompose the time series into its three components. For this purpose, we use the *decompose( )* function defined in the *TTR* library in R. The decomposition results enable us to make a comparative analysis of the behavior of the five time series belonging to five different sectors. We validate several hypotheses by our deeper analysis of the decomposition results.

After a detailed analysis of the decomposition results, we enter into our second endeavor in this work. Based on our deeper understanding about the association among different sectors as observed from their time series analysis, we make bivariate analysis and forecasting using linear regression models. This analysis enables us to forecast the performance of one sector on basis of performance of another to which it is closely associated with. We have also carried out analysis on the forecast accuracies by suitably choosing our training data set for building the linear regression model and test data for testing the effectiveness of our forecasting models.

In our previous work, we have highlighted the effectiveness of time series decomposition approach for robust analysis and forecasting of the Indian Auto sector (Sen & Datta Chaudhuri, 2016a; Sen & Datta Chaudhuri, 2016b) and we have also made a comparative study of the behavioral characteristics of two different sectors of the Indian economy – the Consumer

Durable Goods sector and the Small Cap sector (Sen & Datta Chaudhuri, 2006c). In contrast to our previous work, in this paper, we have presented a detailed study on the structural decomposition of the time series index of Indian IT and CG sectors, NIFTY index, DJIA index and the US Dollar to the Indian Rupee exchange rate. We have then carried out association analyses between these time series to validate several hypotheses that we postulate. Association analyses are carried out using several robust statistical tests. After a comprehensive association analysis, we have constructed several linear regression models between those time series which exhibited strong association among them. To demonstrate the robustness and accuracies of the linear models, we have used the models for forecasting using suitable chosen training and test data sets.

## 3. Time Series Decomposition Results

We now present the methods that we have followed to decompose the time series of five different sectors – Indian IT sector, Indian CG sector, DJIA index, NIFTY index and US Dollar to Indian Rupee exchange rate. For all these sectors, we have first taken the daily index values from January 2009 to April 2016 and saved them in five separate plain text (.txt) files – one file storing the daily time series index of one sector. From these daily index values, we have computed the monthly averages and saved the monthly average values in five separate text files. Each of these text files contained 88 values (records of 7 years and 4 month leading to 88 monthly average values). We used R language function *scan ( )* to read these text files and store them into five appropriate R variables. Then, we converted these five R variables into five time series variables using the R function *ts ( )*, which is defined in the package *TTR*. Once these five time series variables are constructed, we have used the *plot ( )* function in R to derive the displays of the time series for the five sectors under study for the period January 2009 – April 2016. The time series for the Indian IT sector index, the Indian CG sector index, the DJIA index, the NIFTY index and the US Dollar to Indian Rupee exchange rate values are represented in Figure 1, Figure 2, Figure 3, Figure 4, and Figure 5 respectively.

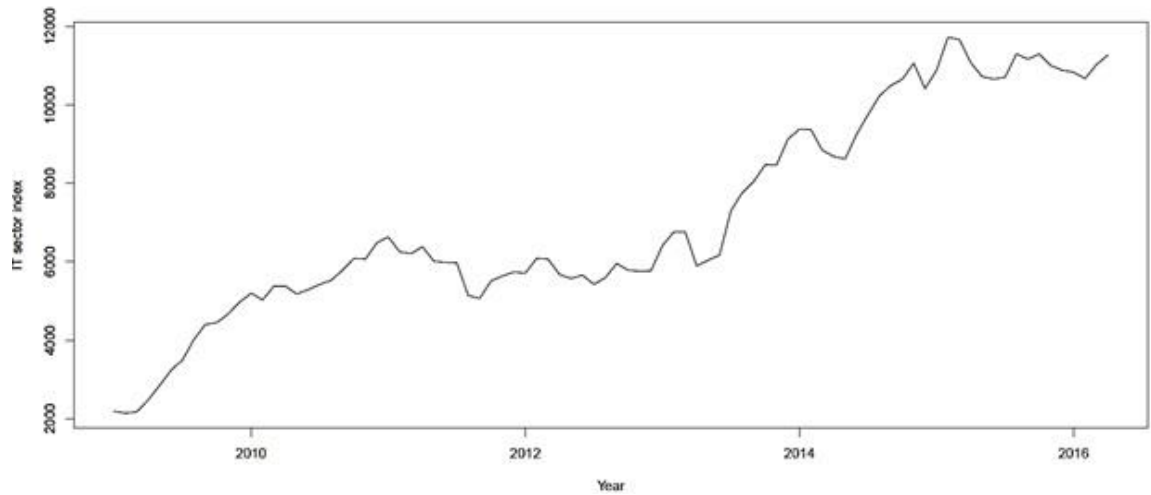

**Figure 1**
The Indian IT sector index time series (Jan 2009 – Apr 2016)

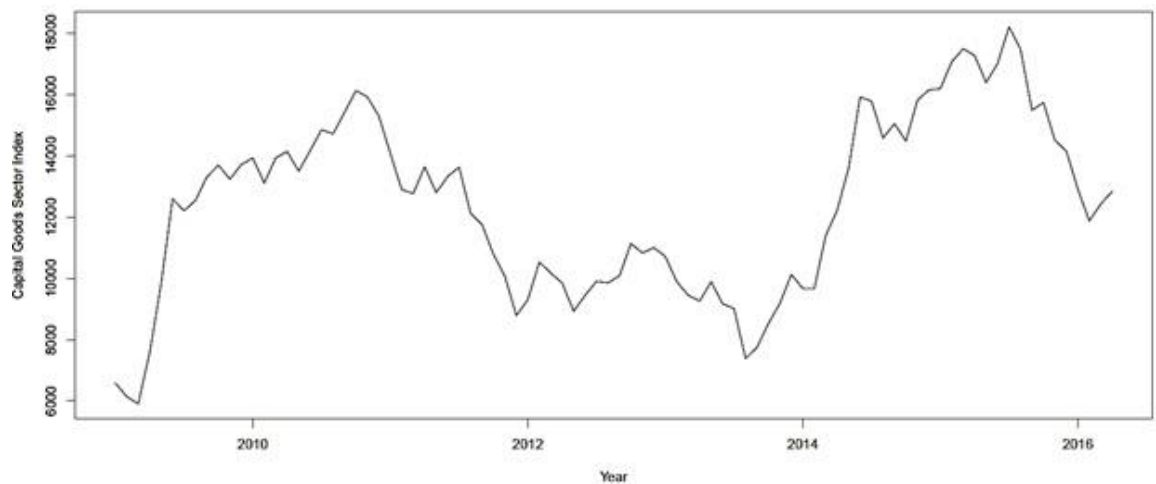

**Figure 2**
The Indian CG sector index time series (Jan 2009 – Apr 2016)

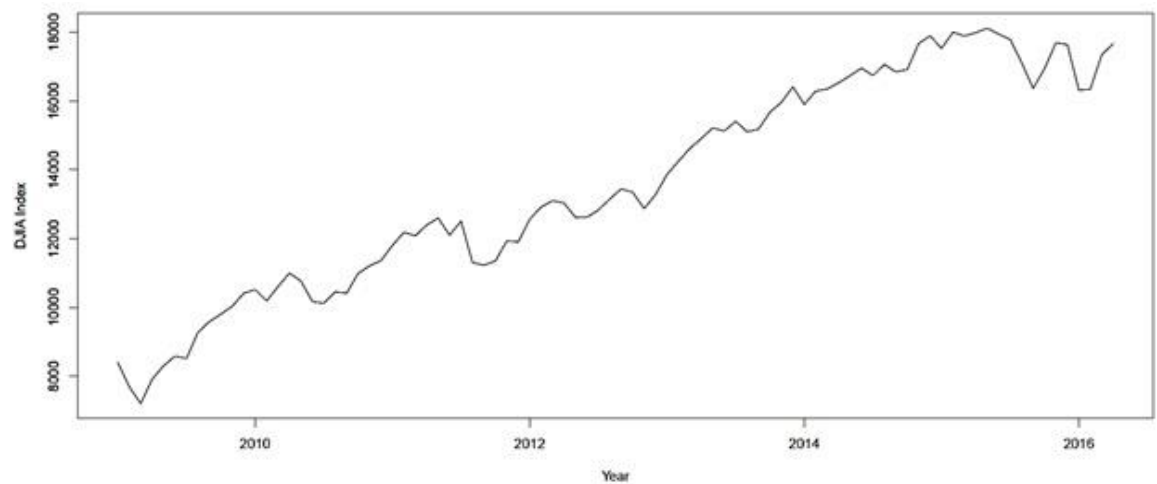

**Figure 3**
The DJIA index time series (Jan 2009 – Apr 2016)

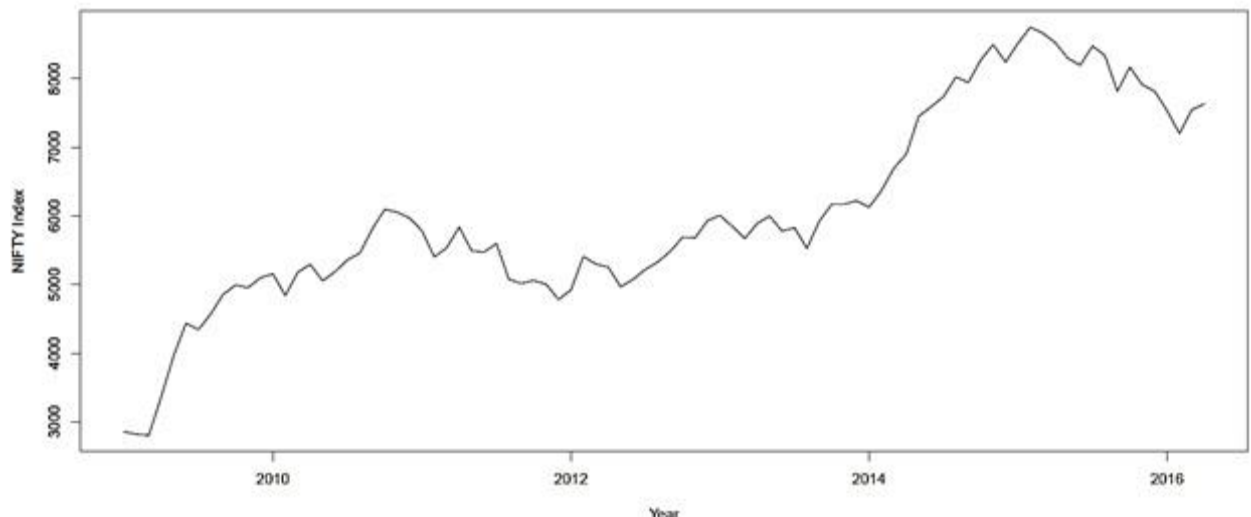

**Figure 4**
The NIFTY index time series (Jan 2009 – Apr 2016)

The plots of the time series for the five sectors exhibit the overall behavior of these time series over the period under consideration (i.e., January 2009 – December 2016). However, to get a deeper insight into these time series, we have decomposed the five time series variables into their trend, seasonal and random components using the *decompose ( )* function that is defined in the TTR library in the R programming environment. Figure 6, Figure 7, Figure 8, Figure 9 and Figure 10 depict the decomposition results for the time series of the Indian IT sector, Indian CG sector, the DJIA index, the NIFTY index and the US Dollar to Indian Rupee exchange rate values. Each of the five figures (Figure 6 to Figure 10) has four boxes arranged in a stack. The boxes depict the overall time series, the trend, the seasonal and the random component respectively, arranged from top to bottom.

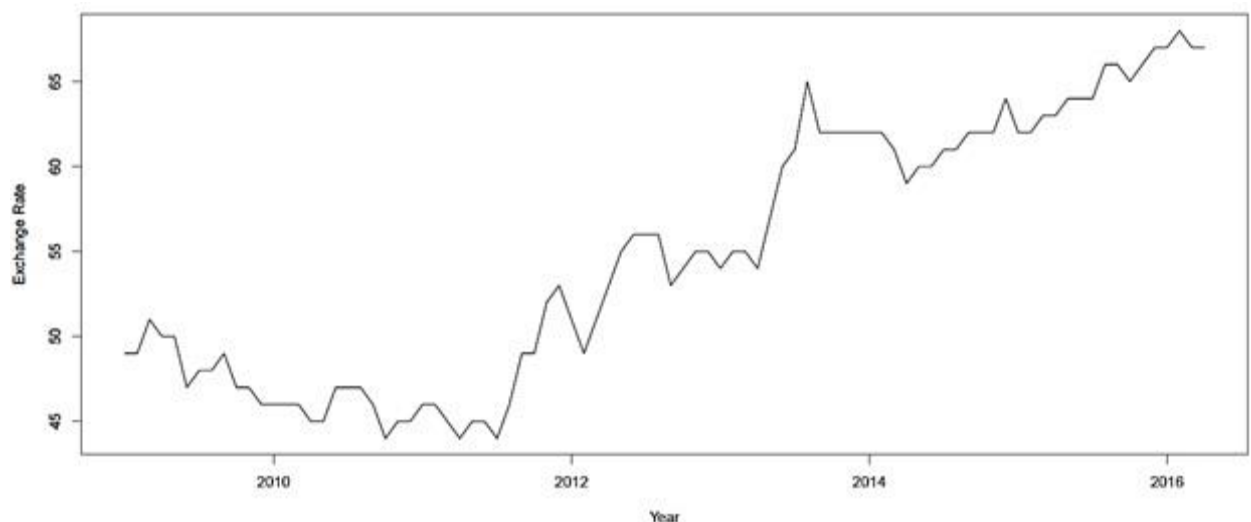

**Figure 5**
The US Dollar to Indian Rupee exchange rate time series (Jan 2009 – Apr 2016)

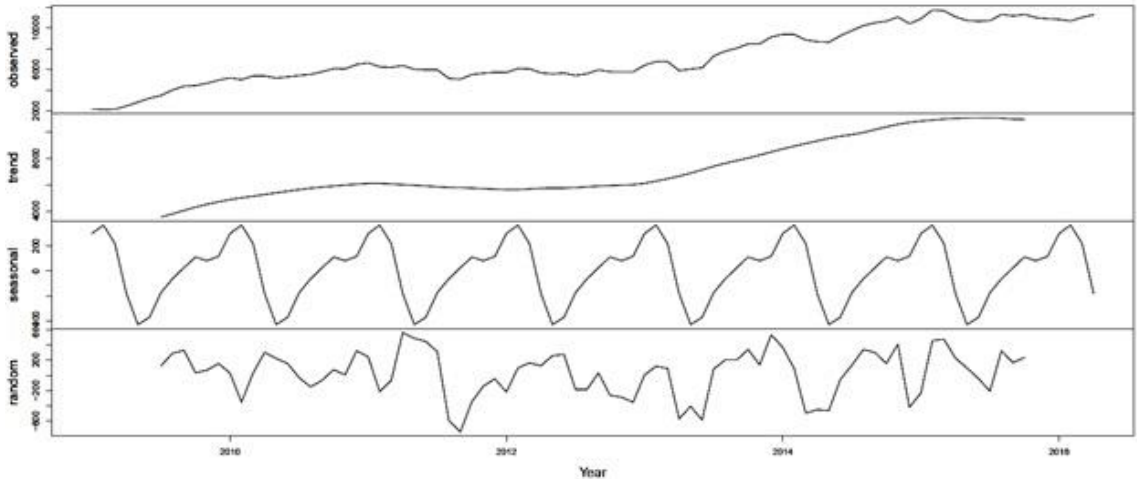

**Figure 6**
Decomposition of Indian IT sector index time series into its trend, seasonal and random components (Jan 2009 – Apr 2016)

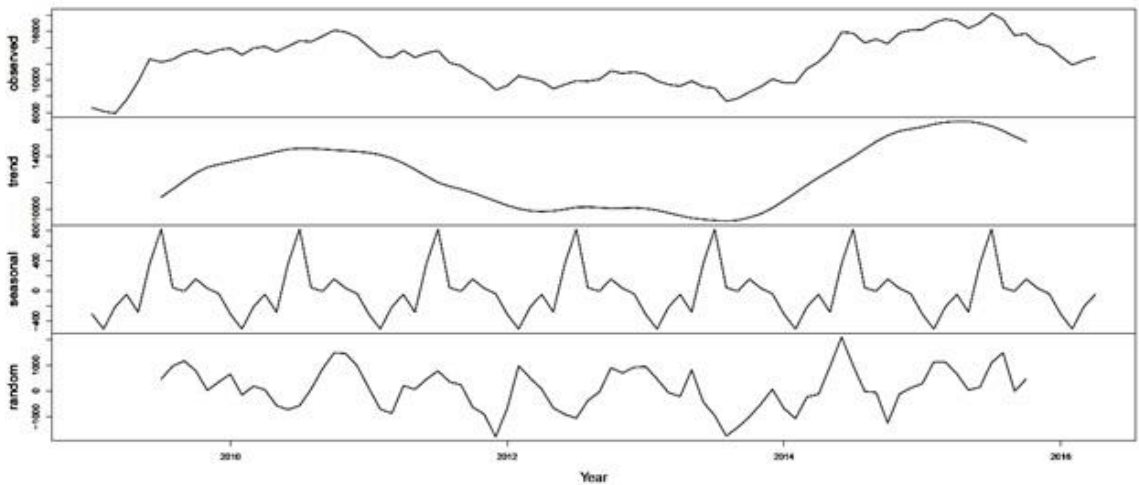

**Figure 7**
Decomposition of Indian CG sector index time series into its trend, seasonal and random components (Jan 2009 – Apr 2016)

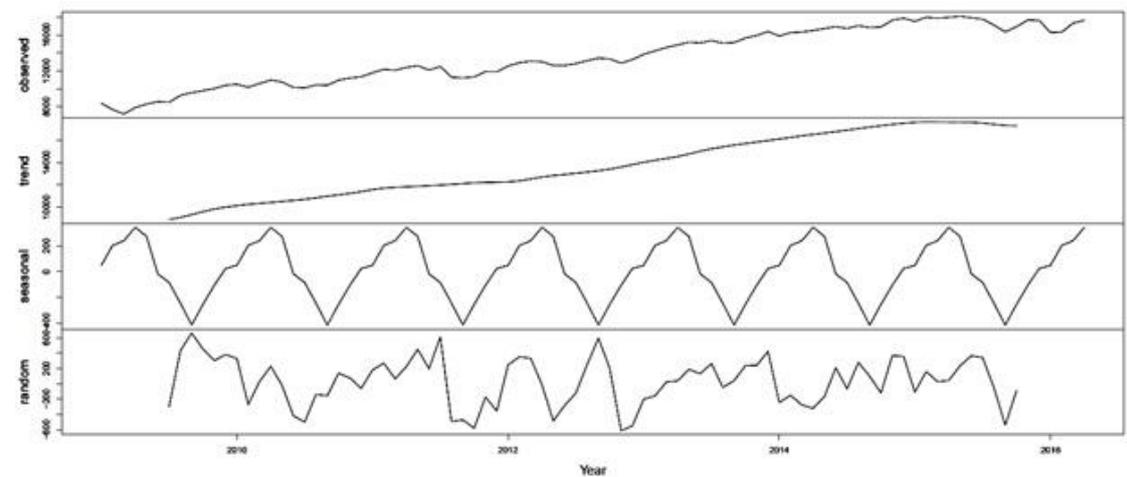

**Figure 8**
Decomposition of DJIA index time series into its trend, seasonal and random components (Jan 2009 – Apr 2016)

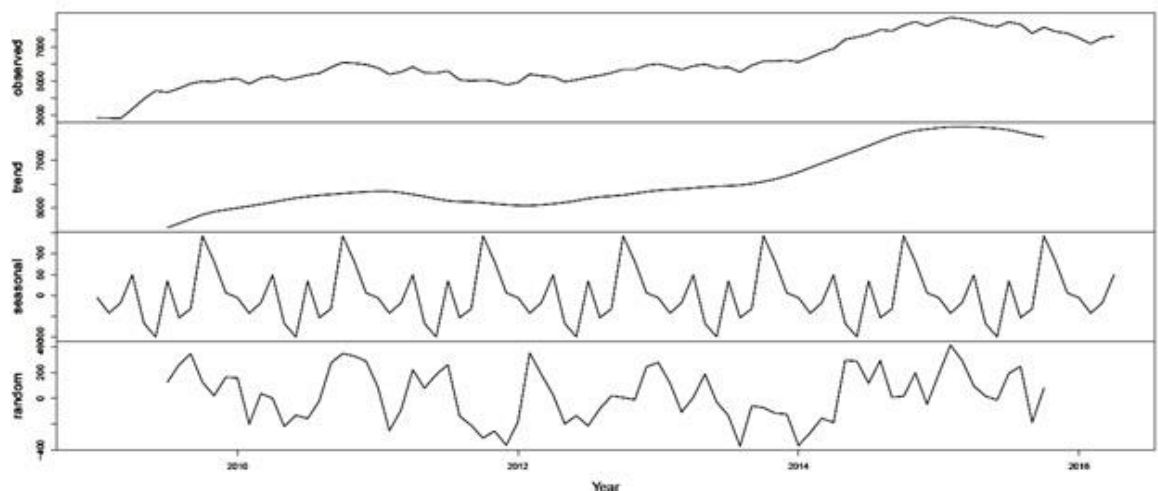

**Figure 9**
Decomposition of the NIFTY index time series into its trend, seasonal and random components (Jan 2009 – Apr 2016)

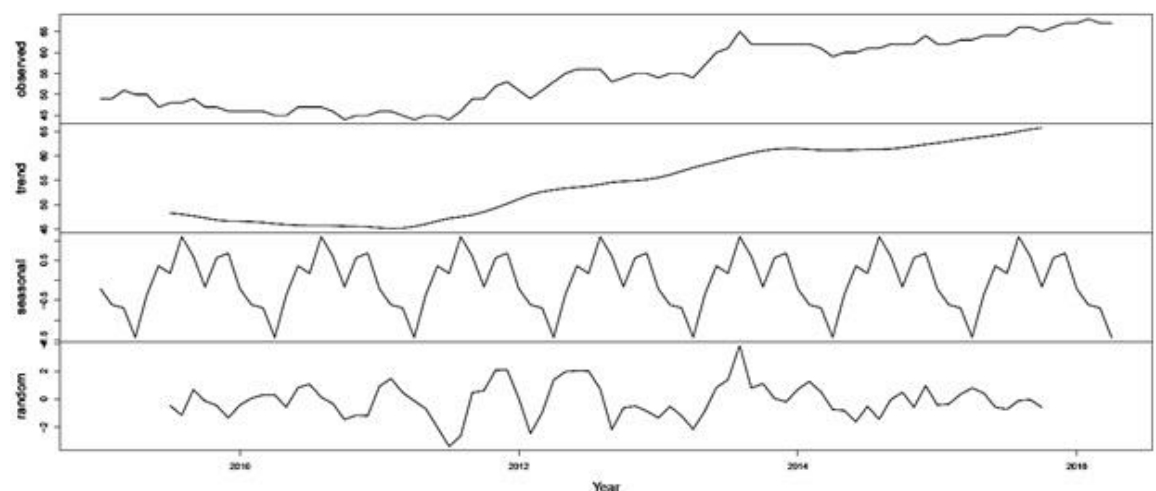

**Figure 10**
Decomposition of the US Dollar to Indian Rupee exchange rate time series into its trend, seasonal and random components (Jan 2009 – Apr 2016)

The numerical values of the time series and its three components for the Indian IT sector, Indian CG sector, the DJIA index, the NIFTY index, and the US Dollar to Indian Rupee exchange rate values are presented in Table 1, Table 2, Table 3, Table 4 and Table 5 respectively. It may be interesting to observe that the values of the trend and the random components are not available for the period January 2009 – June 2009 and also for the period November 2015 – April 2016. Since the *decompose( )* function in R uses a 12 month moving average method for computing the trend component, in order to compute the trend value for January 2009, we need time series data from July 2008 to December 2008. However, since we have used time series data from January 2009 to April 2016, the first trend value the *decompose ( )* function could compute was for the month of July 2009 and the last month being November 2015. For computing the seasonal component, the *decompose ( )* function first *detrends* (subtracts the trend component from the overall time series) the time series and arranges the time series values in a 12 column format. The seasonal values for each month is derived by computing the averages of each column. The value of the seasonal component for a given month remains the same for the entire period under study. The random components are obtained after subtracting the sum of the corresponding trend and seasonal components from the overall time series values. Since the trend values for the period January 2009 – June 2009 and November 2015 – April 2016 are missing, the random components for those periods could not be computed as well.

**Table 1**
Aggregate value of the Indian IT sector index time series and its components
(Jan 2009 – Apr 2016)

| Year | Month | Aggregate | Trend | Seasonal | Random |
|---|---|---|---|---|---|
| 2009 | January | 2189 | | 299 | |
| | February | 2140 | | 367 | |
| | March | 2175 | | 219 | |
| | April | 2483 | | -181 | |
| | May | 2850 | | -429 | |
| | June | 3237 | | -372 | |
| | July | 3500 | 3550 | -175 | 125 |
| | August | 4022 | 3795 | -63 | 290 |
| | September | 4403 | 4049 | 26 | 328 |
| | October | 4449 | 4303 | 111 | 35 |
| | November | 4670 | 4521 | 81 | 68 |
| | December | 4974 | 4703 | 116 | 155 |
| 2010 | January | 5197 | 4869 | 299 | 29 |
| | February | 5026 | 5012 | 367 | -353 |
| | March | 5381 | 5132 | 219 | 30 |
| | April | 5379 | 5258 | -181 | 302 |
| | May | 5177 | 5384 | -429 | 222 |
| | June | 5290 | 5506 | -372 | 156 |
| | July | 5423 | 5628 | -175 | -30 |
| | August | 5525 | 5739 | -63 | -151 |
| | September | 5787 | 5825 | 26 | -64 |
| | October | 6086 | 5901 | 111 | 74 |
| | November | 6067 | 5978 | 81 | 8 |
| | December | 6482 | 6042 | 116 | 324 |
| 2011 | January | 6635 | 6094 | 299 | 242 |
| | February | 6254 | 6101 | 367 | -214 |
| | March | 6207 | 6054 | 219 | -66 |

|      | Month | | | | |
|------|-----------|-------|-------|------|------|
|      | April     | 6382  | 6000  | -181 | 563  |
|      | May       | 6015  | 5958  | -429 | 486  |
|      | June      | 5984  | 5909  | -372 | 445  |
|      | July      | 5975  | 5840  | -175 | 310  |
|      | August    | 5141  | 5794  | -63  | -590 |
|      | September | 5062  | 5782  | 26   | -746 |
|      | October   | 5511  | 5746  | 111  | -346 |
|      | November  | 5637  | 5699  | 81   | -143 |
|      | December  | 5738  | 5667  | 116  | -45  |
| 2012 | January   | 5709  | 5630  | 299  | -220 |
|      | February  | 6089  | 5626  | 367  | 96   |
|      | March     | 6065  | 5682  | 219  | 164  |
|      | April     | 5676  | 5731  | -181 | 126  |
|      | May       | 5575  | 5747  | -429 | 257  |
|      | June      | 5658  | 5754  | -372 | 276  |
|      | July      | 5425  | 5784  | -175 | -184 |
|      | August    | 5592  | 5841  | -63  | -186 |
|      | September | 5957  | 5898  | 26   | 33   |
|      | October   | 5785  | 5936  | 111  | -262 |
|      | November  | 5759  | 5965  | 81   | -287 |
|      | December  | 5768  | 6005  | 116  | -353 |
| 2013 | January   | 6409  | 6105  | 299  | 5    |
|      | February  | 6760  | 6273  | 367  | 120  |
|      | March     | 6761  | 6451  | 219  | 91   |
|      | April     | 5896  | 6650  | -181 | -573 |
|      | May       | 6039  | 6875  | -429 | -407 |
|      | June      | 6168  | 7128  | -372 | -588 |
|      | July      | 7300  | 7392  | -175 | 83   |
|      | August    | 7764  | 7624  | -63  | 203  |
|      | September | 8048  | 7820  | 26   | 202  |
|      | October   | 8473  | 8023  | 111  | 339  |
|      | November  | 8466  | 8247  | 81   | 138  |
|      | December  | 9133  | 8483  | 116  | 534  |
| 2014 | January   | 9379  | 8713  | 299  | 367  |
|      | February  | 9373  | 8918  | 367  | 88   |
|      | March     | 8843  | 9122  | 219  | -498 |
|      | April     | 8684  | 9315  | -181 | -450 |
|      | May       | 8621  | 9513  | -429 | -463 |
|      | June      | 9247  | 9675  | -372 | -56  |
|      | July      | 9748  | 9791  | -175 | 132  |
|      | August    | 10226 | 9952  | -63  | 337  |
|      | September | 10494 | 10167 | 26   | 301  |
|      | October   | 10650 | 10384 | 111  | 155  |

| Year | Month | | | | |
|---|---|---|---|---|---|
| | November | 11061 | 10571 | 81 | 409 |
| | December | 10414 | 10718 | 116 | -420 |
| 2015 | January | 10882 | 10816 | 299 | -233 |
| | February | 11724 | 10901 | 367 | 456 |
| | March | 11667 | 10973 | 219 | 475 |
| | April | 11073 | 11028 | -181 | 226 |
| | May | 10721 | 11053 | -429 | 97 |
| | June | 10656 | 11070 | -372 | -42 |
| | July | 10703 | 11087 | -175 | -209 |
| | August | 11301 | 11041 | -63 | 323 |
| | September | 11164 | 10970 | 26 | 168 |
| | October | 11296 | 10952 | 111 | 233 |
| | November | 10999 | | 81 | |
| | December | 10884 | | 116 | |
| 2016 | January | 10832 | | 299 | |
| | February | 10670 | | 367 | |
| | March | 11023 | | 219 | |
| | April | 11270 | | -181 | |

**Table 2**
Aggregate value of the Indian CG sector index and its components
(Jan 2009 – Apr 2016)

| Year | Month | Aggregate | Trend | Seasonal | Random |
|---|---|---|---|---|---|
| 2009 | January | 6588 | | -309 | |
| | February | 6144 | | -507 | |
| | March | 5906 | | -213 | |
| | April | 7576 | | -47 | |
| | May | 9823 | | -284 | |
| | June | 12597 | | 359 | |
| | July | 12208 | 10918 | 817 | 474 |
| | August | 12540 | 11514 | 41 | 986 |
| | September | 13315 | 12138 | -6 | 1183 |
| | October | 13698 | 12746 | 156 | 796 |
| | November | 13234 | 13173 | 33 | 28 |
| | December | 13714 | 13391 | -39 | 362 |
| 2010 | January | 13926 | 13566 | -309 | 669 |
| | February | 13111 | 13767 | -507 | -149 |
| | March | 13927 | 13945 | -213 | 196 |
| | April | 14140 | 14133 | -47 | 54 |
| | May | 13497 | 14346 | -284 | -565 |
| | June | 14163 | 14524 | 359 | -720 |
| | July | 14850 | 14596 | 817 | -563 |
| | August | 14712 | 14594 | 41 | 78 |
| | September | 15411 | 14536 | -6 | 880 |
| | October | 16126 | 14467 | 156 | 1503 |
| | November | 15919 | 14417 | 33 | 1469 |
| | December | 15296 | 14353 | -39 | 982 |
| 2011 | January | 14079 | 14267 | -309 | 121 |
| | February | 12898 | 14109 | -507 | -704 |

|      | Month     |       |       |      |       |
|------|-----------|-------|-------|------|-------|
|      | March     | 12766 | 13849 | -213 | -869  |
|      | April     | 13641 | 13474 | -47  | 214   |
|      | May       | 12795 | 13008 | -284 | 71    |
|      | June      | 13320 | 12493 | 359  | 468   |
|      | July      | 13633 | 12023 | 817  | 793   |
|      | August    | 12128 | 11726 | 41   | 362   |
|      | September | 11762 | 11519 | -6   | 249   |
|      | October   | 10786 | 11253 | 156  | -623  |
|      | November  | 10070 | 10934 | 33   | -897  |
|      | December  | 8786  | 10611 | -39  | -1787 |
| 2012 | January   | 9306  | 10295 | -309 | -679  |
|      | February  | 10530 | 10045 | -507 | 992   |
|      | March     | 10174 | 9880  | -213 | 507   |
|      | April     | 9851  | 9825  | -47  | 73    |
|      | May       | 8928  | 9871  | -284 | -659  |
|      | June      | 9447  | 9995  | 359  | -907  |
|      | July      | 9907  | 10146 | 817  | -1056 |
|      | August    | 9855  | 10179 | 41   | -364  |
|      | September | 10085 | 10122 | -6   | -31   |
|      | October   | 11136 | 10067 | 156  | 913   |
|      | November  | 10829 | 10083 | 33   | 713   |
|      | December  | 11003 | 10112 | -39  | 930   |
| 2013 | January   | 10717 | 10063 | -309 | 963   |
|      | February  | 9894  | 9923  | -507 | 477   |
|      | March     | 9451  | 9723  | -213 | -59   |
|      | April     | 9262  | 9518  | -47  | -208  |
|      | May       | 9890  | 9341  | -284 | 834   |
|      | June      | 9172  | 9236  | 359  | -423  |
|      | July      | 9018  | 9156  | 817  | -954  |
|      | August    | 7390  | 9102  | 41   | -1753 |
|      | September | 7753  | 9174  | -6   | -1415 |
|      | October   | 8528  | 9378  | 156  | -1006 |
|      | November  | 9188  | 9657  | 33   | -502  |
|      | December  | 10128 | 10092 | -39  | 74    |
| 2014 | January   | 9669  | 10655 | -309 | -677  |
|      | February  | 9661  | 11237 | -507 | -1069 |
|      | March     | 11400 | 11840 | -213 | -226  |
|      | April     | 12225 | 12392 | -47  | -119  |
|      | May       | 13603 | 12915 | -284 | 972   |
|      | June      | 15918 | 13442 | 359  | 2117  |
|      | July      | 15782 | 13965 | 817  | 1001  |
|      | August    | 14577 | 14545 | 41   | -8    |
|      | September | 15048 | 15107 | -6   | -53   |
|      | October   | 14475 | 15571 | 156  | -1252 |
|      | November  | 15810 | 15897 | 33   | -121  |
|      | December  | 16149 | 16059 | -39  | 129   |
| 2015 | January   | 16189 | 16206 | -309 | 293   |
|      | February  | 17064 | 16427 | -507 | 1143  |
|      | March     | 17495 | 16566 | -213 | 1142  |
|      | April     | 17266 | 16637 | -47  | 677   |
|      | May       | 16389 | 16635 | -284 | 38    |
|      | June      | 17013 | 16497 | 359  | 157   |
|      | July      | 18206 | 16278 | 817  | 1112  |
|      | August    | 17471 | 15925 | 41   | 1506  |

|  | September | 15485 | 15497 | -6 | -7 |
|---|---|---|---|---|---|
|  | October | 15734 | 15102 | 156 | 476 |
|  | November | 14501 |  | 33 |  |
|  | December | 14155 |  | -39 |  |
| **2016** | January | 12914 |  | -309 |  |
|  | February | 11875 |  | -507 |  |
|  | March | 12424 |  | -213 |  |
|  | April | 12837 |  | -47 |  |

**Table 3**
Aggregate value of the DJIA index and its components (Jan 2009 – Apr 2016)

| Year | Month | Aggregate | Trend | Seasonal | Random |
|---|---|---|---|---|---|
| **2009** | January | 8396 |  | 48 |  |
|  | February | 7690 |  | 203 |  |
|  | March | 7202 |  | 239 |  |
|  | April | 7914 |  | 343 |  |
|  | May | 8302 |  | 273 |  |
|  | June | 8581 |  | -19 |  |
|  | July | 8516 | 8897 | -85 | -296 |
|  | August | 9275 | 9090 | -247 | 432 |
|  | September | 9584 | 9335 | -417 | 665 |
|  | October | 9802 | 9606 | -255 | 452 |
|  | November | 10033 | 9837 | -107 | 303 |
|  | December | 10412 | 10006 | 23 | 383 |
| **2010** | January | 10516 | 10139 | 48 | 330 |
|  | February | 10186 | 10254 | 203 | -272 |
|  | March | 10606 | 10338 | 239 | 29 |
|  | April | 10995 | 10422 | 343 | 230 |
|  | May | 10769 | 10520 | 273 | -24 |
|  | June | 10170 | 10609 | -19 | -420 |
|  | July | 10116 | 10702 | -85 | -500 |
|  | August | 10454 | 10838 | -247 | -138 |
|  | September | 10411 | 10983 | -417 | -156 |
|  | October | 10987 | 11103 | -255 | 139 |
|  | November | 11207 | 11238 | -107 | 76 |
|  | December | 11359 | 11395 | 23 | -59 |
| **2011** | January | 11802 | 11575 | 48 | 180 |
|  | February | 12183 | 11710 | 203 | 270 |
|  | March | 12084 | 11779 | 239 | 66 |
|  | April | 12396 | 11828 | 343 | 224 |
|  | May | 12599 | 11874 | 273 | 452 |
|  | June | 12104 | 11927 | -19 | 196 |
|  | July | 12508 | 11981 | -85 | 612 |
|  | August | 11302 | 12043 | -247 | -494 |
|  | September | 11228 | 12116 | -417 | -472 |
|      | October   | 11349 | 12185 | -255 | -581 |
|      | November  | 11932 | 12213 | -107 | -174 |
|      | December  | 11903 | 12235 | 23   | -355 |
| 2012 | January   | 12565 | 12270 | 48   | 248  |
|      | February  | 12916 | 12360 | 203  | 353  |
|      | March     | 13102 | 12530 | 239  | 333  |
|      | April     | 13035 | 12705 | 343  | -14  |
|      | May       | 12615 | 12828 | 273  | -486 |
|      | June      | 12621 | 12925 | -19  | -285 |
|      | July      | 12829 | 13036 | -85  | -122 |
|      | August    | 13152 | 13146 | -247 | 253  |
|      | September | 13446 | 13265 | -417 | 598  |
|      | October   | 13349 | 13406 | -255 | 198  |
|      | November  | 12873 | 13593 | -107 | -613 |
|      | December  | 13279 | 13806 | 23   | -550 |
| 2013 | January   | 13864 | 14018 | 48   | -201 |
|      | February  | 14251 | 14207 | 203  | -159 |
|      | March     | 14622 | 14361 | 239  | 22   |
|      | April     | 14912 | 14530 | 343  | 38   |
|      | May       | 15219 | 14756 | 273  | 190  |
|      | June      | 15128 | 15015 | -19  | 131  |
|      | July      | 15410 | 15231 | -85  | 265  |
|      | August    | 15108 | 15401 | -247 | -46  |
|      | September | 15181 | 15558 | -417 | 40   |
|      | October   | 15679 | 15697 | -255 | 237  |
|      | November  | 15961 | 15828 | -107 | 240  |
|      | December  | 16415 | 15967 | 23   | 424  |
| 2014 | January   | 15902 | 16099 | 48   | -245 |
|      | February  | 16289 | 16236 | 203  | -150 |
|      | March     | 16350 | 16387 | 239  | -276 |
|      | April     | 16531 | 16509 | 343  | -321 |
|      | May       | 16737 | 16632 | 273  | -167 |
|      | June      | 16958 | 16765 | -19  | 212  |
|      | July      | 16743 | 16894 | -85  | -66  |
|      | August    | 17066 | 17034 | -247 | 279  |
|      | September | 16852 | 17169 | -417 | 99   |
|      | October   | 16918 | 17294 | -255 | -121 |
|      | November  | 17674 | 17413 | -107 | 368  |
|      | December  | 17893 | 17511 | 23   | 358  |
| 2015 | January   | 17534 | 17596 | 48   | -110 |
|      | February  | 18005 | 17642 | 203  | 160  |
|      | March     | 17891 | 17624 | 239  | 28   |
|      | April     | 17992 | 17605 | 343  | 44   |

|  | Month | | | |
|---|---|---|---|---|
|  | May | 18116 | 17607 | 273 | 236 |
|  | June | 17945 | 17597 | -19 | 366 |
|  | July | 17792 | 17536 | -85 | 342 |
|  | August | 17117 | 17416 | -247 | -52 |
|  | September | 16367 | 17324 | -417 | -540 |
|  | October | 16944 | 17287 | -255 | -88 |
|  | November | 17697 |  | -107 |  |
|  | December | 17639 |  | 23 |  |
| **2016** | January | 16312 |  | 48 |  |
|  | February | 16348 |  | 203 |  |
|  | March | 17344 |  | 239 |  |
|  | April | 17665 |  | 343 |  |

**Table 4**
Aggregate value of the NIFTY index and its components (Jan 2009 – Apr 2016)

| Year | Month | Aggregate | Trend | Seasonal | Random |
|---|---|---|---|---|---|
| **2009** | January | 2854 |  | -5 |  |
|  | February | 2819 |  | -42 |  |
|  | March | 2802 |  | -17 |  |
|  | April | 3360 |  | 50 |  |
|  | May | 3958 |  | -67 |  |
|  | June | 4436 |  | -99 |  |
|  | July | 4343 | 4183 | 35 | 124 |
|  | August | 4571 | 4364 | -53 | 261 |
|  | September | 4859 | 4547 | -32 | 344 |
|  | October | 4994 | 4726 | 143 | 125 |
|  | November | 4954 | 4853 | 81 | 20 |
|  | December | 5100 | 4930 | 6 | 164 |
| **2010** | January | 5156 | 5003 | -5 | 158 |
|  | February | 4840 | 5083 | -42 | -200 |
|  | March | 5178 | 5159 | -17 | 36 |
|  | April | 5295 | 5245 | 50 | 1 |
|  | May | 5053 | 5337 | -67 | -217 |
|  | June | 5188 | 5419 | -99 | -132 |
|  | July | 5360 | 5481 | 35 | -156 |
|  | August | 5457 | 5531 | -53 | -20 |
|  | September | 5811 | 5569 | -32 | 274 |
|  | October | 6096 | 5607 | 143 | 346 |
|  | November | 6055 | 5648 | 81 | 326 |
|  | December | 5971 | 5678 | 6 | 287 |
| **2011** | January | 5783 | 5700 | -5 | 89 |
|  | February | 5401 | 5694 | -42 | -250 |

|  | March | 5538 | 5645 | -17 | -89 |
|---|---|---|---|---|---|
|  | April | 5839 | 5568 | 50 | 221 |
|  | May | 5492 | 5481 | -67 | 77 |
|  | June | 5473 | 5388 | -99 | 184 |
|  | July | 5597 | 5303 | 35 | 259 |
|  | August | 5077 | 5267 | -53 | -137 |
|  | September | 5016 | 5257 | -32 | -210 |
|  | October | 5060 | 5223 | 143 | -306 |
|  | November | 5004 | 5177 | 81 | -254 |
|  | December | 4782 | 5138 | 6 | -363 |
| **2012** | January | 4920 | 5106 | -5 | -181 |
|  | February | 5409 | 5101 | -42 | 350 |
|  | March | 5298 | 5131 | -17 | 184 |
|  | April | 5254 | 5177 | 50 | 28 |
|  | May | 4967 | 5231 | -67 | -197 |
|  | June | 5074 | 5307 | -99 | -134 |
|  | July | 5222 | 5400 | 35 | -213 |
|  | August | 5330 | 5464 | -53 | -81 |
|  | September | 5485 | 5498 | -32 | 19 |
|  | October | 5689 | 5540 | 143 | 6 |
|  | November | 5680 | 5609 | 81 | -10 |
|  | December | 5932 | 5682 | 6 | 244 |
| **2013** | January | 6008 | 5736 | -5 | 277 |
|  | February | 5846 | 5770 | -42 | 119 |
|  | March | 5672 | 5796 | -17 | -107 |
|  | April | 5891 | 5835 | 50 | 7 |
|  | May | 5997 | 5875 | -67 | 189 |
|  | June | 5779 | 5908 | -99 | -29 |
|  | July | 5827 | 5925 | 35 | -133 |
|  | August | 5528 | 5951 | -53 | -370 |
|  | September | 5925 | 6016 | -32 | -59 |
|  | October | 6171 | 6100 | 143 | -72 |
|  | November | 6169 | 6203 | 81 | -115 |
|  | December | 6223 | 6339 | 6 | -122 |
| **2014** | January | 6124 | 6494 | -5 | -365 |
|  | February | 6369 | 6678 | -42 | -266 |
|  | March | 6695 | 6866 | -17 | -154 |
|  | April | 6899 | 7038 | 50 | -188 |
|  | May | 7450 | 7222 | -67 | 294 |
|  | June | 7591 | 7403 | -99 | 287 |
|  | July | 7739 | 7587 | 35 | 117 |
|  | August | 8025 | 7786 | -53 | 292 |
|  | September | 7944 | 7967 | -32 | 8 |
|      | October   | 8276 | 8117 | 143 | 16   |
|      | November  | 8497 | 8220 | 81  | 196  |
|      | December  | 8241 | 8281 | 6   | -46  |
| 2015 | January   | 8518 | 8337 | -5  | 186  |
|      | February  | 8750 | 8381 | -42 | 412  |
|      | March     | 8664 | 8388 | -17 | 293  |
|      | April     | 8524 | 8379 | 50  | 96   |
|      | May       | 8300 | 8350 | -67 | 17   |
|      | June      | 8196 | 8308 | -99 | -13  |
|      | July      | 8477 | 8249 | 35  | 193  |
|      | August    | 8337 | 8144 | -53 | 246  |
|      | September | 7816 | 8033 | -32 | -185 |
|      | October   | 8169 | 7949 | 143 | 77   |
|      | November  | 7913 |      | 81  |      |
|      | December  | 7818 |      | 6   |      |
| 2016 | January   | 7536 |      | -5  |      |
|      | February  | 7200 |      | -42 |      |
|      | March     | 7550 |      | -17 |      |
|      | April     | 7632 |      | 50  |      |

**Table 5**
Aggregate value of the US Dollar to Indian Rupees exchange rate time series and its components (Jan 2009 – Apr 2016)

| Year | Month     | Aggregate | Trend | Seasonal | Random |
|------|-----------|-----------|-------|----------|--------|
| 2009 | January   | 49        |       | -0.2     |        |
|      | February  | 49        |       | -0.6     |        |
|      | March     | 51        |       | -0.7     |        |
|      | April     | 50        |       | -1.4     |        |
|      | May       | 50        |       | -0.4     |        |
|      | June      | 47        |       | 0.4      |        |
|      | July      | 48        | 48.3  | 0.2      | -0.4   |
|      | August    | 48        | 48    | 1.1      | -1.1   |
|      | September | 49        | 47.7  | 0.6      | 0.7    |
|      | October   | 47        | 47.3  | -0.2     | -0.1   |
|      | November  | 47        | 46.9  | 0.6      | -0.4   |
|      | December  | 46        | 46.7  | 0.7      | -1.3   |
| 2010 | January   | 46        | 46.6  | -0.2     | -0.4   |
|      | February  | 46        | 46.5  | -0.6     | 0.1    |
|      | March     | 46        | 46.4  | -0.7     | 0.3    |
|      | April     | 45        | 46.1  | -1.4     | 0.3    |
|      | May       | 45        | 46    | -0.4     | -0.6   |
|      | June      | 47        | 45.8  | 0.4      | 0.8    |
|      | July      | 47        | 45.8  | 0.2      | 1.1    |

|      | August    | 47 | 45.8 | 1.1  | 0.2  |
|------|-----------|----|------|------|------|
|      | September | 46 | 45.7 | 0.6  | -0.3 |
|      | October   | 44 | 45.6 | -0.2 | -1.5 |
|      | November  | 45 | 45.6 | 0.6  | -1.1 |
|      | December  | 45 | 45.5 | 0.7  | -1.2 |
| 2011 | January   | 46 | 45.3 | -0.2 | 0.9  |
|      | February  | 46 | 45.1 | -0.6 | 1.5  |
|      | March     | 45 | 45.2 | -0.7 | 0.5  |
|      | April     | 44 | 45.5 | -1.4 | -0.1 |
|      | May       | 45 | 46   | -0.4 | -0.7 |
|      | June      | 45 | 46.7 | 0.4  | -2   |
|      | July      | 44 | 47.2 | 0.2  | -3.4 |
|      | August    | 46 | 47.5 | 1.1  | -2.6 |
|      | September | 49 | 47.9 | 0.6  | 0.5  |
|      | October   | 49 | 48.5 | -0.2 | 0.6  |
|      | November  | 52 | 49.3 | 0.6  | 2.1  |
|      | December  | 53 | 50.2 | 0.7  | 2.1  |
| 2012 | January   | 51 | 51.2 | -0.2 | 0    |
|      | February  | 49 | 52.1 | -0.6 | -2.5 |
|      | March     | 51 | 52.7 | -0.7 | -1   |
|      | April     | 53 | 53   | -1.4 | 1.4  |
|      | May       | 55 | 53.4 | -0.4 | 2    |
|      | June      | 56 | 53.6 | 0.4  | 2.1  |
|      | July      | 56 | 53.8 | 0.2  | 2    |
|      | August    | 56 | 54.2 | 1.1  | 0.7  |
|      | September | 53 | 54.6 | 0.6  | -2.2 |
|      | October   | 54 | 54.8 | -0.2 | -0.6 |
|      | November  | 55 | 54.9 | 0.6  | -0.5 |
|      | December  | 55 | 55.2 | 0.7  | -0.8 |
| 2013 | January   | 54 | 55.5 | -0.2 | -1.3 |
|      | February  | 55 | 56.1 | -0.6 | -0.5 |
|      | March     | 55 | 56.9 | -0.7 | -1.1 |
|      | April     | 54 | 57.6 | -1.4 | -2.1 |
|      | May       | 57 | 58.2 | -0.4 | -0.8 |
|      | June      | 60 | 58.8 | 0.4  | 0.8  |
|      | July      | 61 | 59.4 | 0.2  | 1.4  |
|      | August    | 65 | 60   | 1.1  | 3.9  |
|      | September | 62 | 60.6 | 0.6  | 0.8  |
|      | October   | 62 | 61   | -0.2 | 1.1  |
|      | November  | 62 | 61.4 | 0.6  | 0.1  |
|      | December  | 62 | 61.5 | 0.7  | -0.2 |
| 2014 | January   | 62 | 61.5 | -0.2 | 0.7  |
|      | February  | 62 | 61.3 | -0.6 | 1.3  |
|      | March     | 61 | 61.2 | -0.7 | 0.5  |
|      | April     | 59 | 61.2 | -1.4 | -0.7 |
|      | May       | 60 | 61.2 | -0.4 | -0.8 |
|      | June      | 60 | 61.3 | 0.4  | -1.6 |
|      | July      | 61 | 61.3 | 0.2  | -0.5 |
|      | August    | 61 | 61.3 | 1.1  | -1.4 |
|      | September | 62 | 61.4 | 0.6  | 0    |
|      | October   | 62 | 61.7 | -0.2 | 0.5  |
|      | November  | 62 | 62   | 0.6  | -0.6 |
|      | December  | 64 | 62.3 | 0.7  | 1    |
| 2015 | January   | 62 | 62.6 | -0.2 | -0.4 |
|      | February  | 62 | 62.9 | -0.6 | -0.3 |
|      | March     | 63 | 63.3 | -0.7 | 0.4  |
|      | April     | 63 | 63.6 | -1.4 | 0.8  |
|      | May       | 64 | 64   | -0.4 | 0.4  |
|      | June      | 64 | 64.2 | 0.4  | -0.6 |
|      | July      | 64 | 64.5 | 0.2  | -0.7 |
|      | August    | 66 | 65   | 1.1  | -0.1 |
|      | September | 66 | 65.4 | 0.6  | 0    |
|      | October   | 65 | 65.8 | -0.2 | -0.6 |
|      | November  | 66 |      | 0.6  |      |
|      | December  | 67 |      | 0.7  |      |
| 2016 | January   | 67 |      | -0.2 |      |
|      | February  | 68 |      | -0.6 |      |
|      | March     | 67 |      | -0.7 |      |
|      | April     | 67 |      | -1.4 |      |

### 3.1 Analysis of the Time Series Decomposition Results

In this Section, we make a brief analysis of the behavior of each of the five time series and its constituent components. Results of more detailed investigation and analysis have been presented in Section 4.

**Indian IT sector time series:** The Indian IT sector time series in Figure 1 depicts that from January 2009 till July 2011 the time series experienced a modest rate of growth. However, from August 2011 till November 2013 the time series had been rather sluggish with occasional decrease in its values. From December 2013 to October 2015 the IT sector time series index have consistently increased again, before experiencing stagnation again from December 2015 till April 2016. The behavior of the trend component of the IT sector time series can be observed from Figure 6 and Table 1. The trend increased at a slow rate from January 2009 till February 2011. However, the trend started experiencing a fall from March 2011 and the downward trend continued till February 2012 before it started increasing again from March 2012. The trend increased at a very slow rate till February 2013 before it picked up a faster rate of increase from March 2013. However, the trend again started stagnating from March 2015 which continued till October 2015, which is the last trend figure that we could obtain in our study. We can also see the behavior of the seasonal component of the IT sector time series in Figure 6 and Table 1. It is observed that IT sector has a positive seasonal effect during September to March, while the seasonality impact is negative during April to July. The month of February has the highest positive seasonality in the time series, while the month of May has the most negative seasonality component. It is also observed that both the seasonal and the random components have very less magnitudes as compared to the trend component in the time series.

**Indian CG sector time series:** The Indian CG sector time series experienced quite a large number of trend reversals as can be observed from Figure 7 and Table 2. However, roughly, we can divide the time series in four broad time horizons – (i) January 2009 – October 2010, a period during which the CG sector has experienced an upward movement, (ii) November 2010 – September 2013, when the sector has undergone a fall, (iii) October 2013 – July 2015, a period during which the sector had a rise again, and (iv) August 2015 – April 2016, when the sector witnessed a fall again which continued till the end of the time horizon under our study. The trend component of the time series also followed the same pattern. From Table 2, it may be easily seen that the CG sector has a positive seasonality effect during the months of June, July and August with the highest positive seasonality being found in the month of July. The seasonality is negative during the months of January to May, with the most negative value occurring in the month of February. The random component, in general, has more dominant presence than the seasonal component. However, as in the IT time series, the trend is the most predominant component in the CG time series.

**DJIA index time series:** It is evident from Figure 3 and Table 3 that the DJIA time series index consistently increased during the entire period of our study, i.e., January 2009 – April 2016. A careful look at the trend component in Figure 7 makes it evident that the trend stagnated from January 2015 till October 2015 – the last month for which the trend values could be computed. From Table 3, it is also clear that DJIA index have positive seasonality during January to May.

However, the months of June to November experience negative seasonal effects for DJIA index. The random component values are usually higher than those of the seasonal components. However, as in the India IT and Indian CG sector time series, the trend is the most dominant component in the DJIA index time series.

**NIFTY index time series:** As it can be observed from Figure 4, the NIFTY time series has a number spikes and falls. However, we can broadly divide the time horizon into four divisions based on the behavior of the time series: (i) from January 2009 till October 2010, the NIFTY index had an overall upward movement, (ii) during November 2010 to August 2013 there was no substantial change in the NIFTY index values, (iii) from September 2013 to February 2015, the NIFTY index had again increased consistently, and (iv) during March 2015 to April 2016, the NIFTY index experienced a consistent fall. The trend component of the NIFTY time series exhibited the same behavior as can be observed from Figure 7. From Table 4, it is easy to observe that the seasonal component values are very nominal for the NIFTY time series. The month of October experiences the highest positive seasonality while the maximum negative seasonality is observed during the month of June. It is also clear from Table 4 that the random component values are more dominant than those of the seasonal component, while trend is the strongest component in the aggregate time series of NIFTY.

**US Dollar to Indian Rupee exchange rate time series:** Figure 5 depicts the time series for the US Dollar to Indian Rupee exchange rates for the period January 2009 to April 2016. Again, based on the behavior of the time series, the time horizon can be divided into four intervals: (i) from January 2009 to August 2011, during which the exchange rate exhibited a slight downward trend, (ii) from September 2011 to May 2012, the period that experienced a moderate increase in the exchange rate, (iii) from June 2012 to May 2013, during which the exchange rate almost remained constant, and (iv) from June 2013 to April 2016, a period during which the exchange rate increased consistently. From Table 5 and Figure 10, it is evident that the trend of the time series also exhibited similar behavior. The seasonal and the random components in the time series are found to have negligible values compared to those of the trend signifying that the time series is predominantly composed of the trend component only.

## 4. Association Analysis of the Time Series

In order to investigate further into the behavior of the five time series, we carry out several experiments. In this Section, we discuss the details of the studies that we carried out and present the results obtained. The experiments that we have carried out can be broadly categorized into two groups: (i) association analysis of the Indian IT sector time series with the time series of DJIA, NIFT and the Dollar to Rupee exchange rate, (ii) association analysis of the Indian CG sector time series with the time series of DJIA, NIFT and the US Dollar to Indian Rupee exchange rate. This is driven by our two hypotheses: (i) The Indian IT sector is dependent on the overall world economy, and hence the IT time series is expected to be strongly coupled with the DJIA and the Dollar to Rupee exchange rate time series. IT time series is also expected to be strongly associated with the NIFTY time series since the stock prices of some of the blue chip IT companies (e.g., Tata Consultancy Services, Infosys Ltd etc.) have strong impacts on the NIFTY index values. (ii) The CG sector of India is based on India's growth story and hence the CG sector time series is expected to have a strong association with the NIFTY index values. Since the DJIA index and the US Dollar to Indian Rupee exchange rate are related to the world economy, the CG sector time series of India is expected to have a very less association with these two time series.

In order to verify the above two hypotheses, we carry out bivariate correlation tests and cross correlation tests (Shumway & Stoffer, 2011) of the both the IT time series and the CG time series with the DJIA, NIFTY and the US Dollar to Indian Rupee exchange rate time series. In Section 4.1 and Section 4.2, we present the detailed results of the studies of the Indian IT sector and the Indian CG sector with respect to their associations with the DJIA, NIFTY and the Dollar to Rupee exchange rates.

### 4.1 Association Analysis of the Indian IT Time Series

We have tested association of the IT time series with each of the time series of DJIA, NIFTY and Dollar to Rupees exchange rate time series. The detailed results of the study are discussed in this Section.

**4.1.1 Association between the Indian IT and DJIA time series**

In order to study the association between the Indian IT sector time series and the DJIA index time series, we have first plotted the aggregate time series of both the sectors for the period January 2009 to April 2016. Figure 11 depicts the plot. It can be easily observed that the two time series exhibited similar behavior during the period under study.

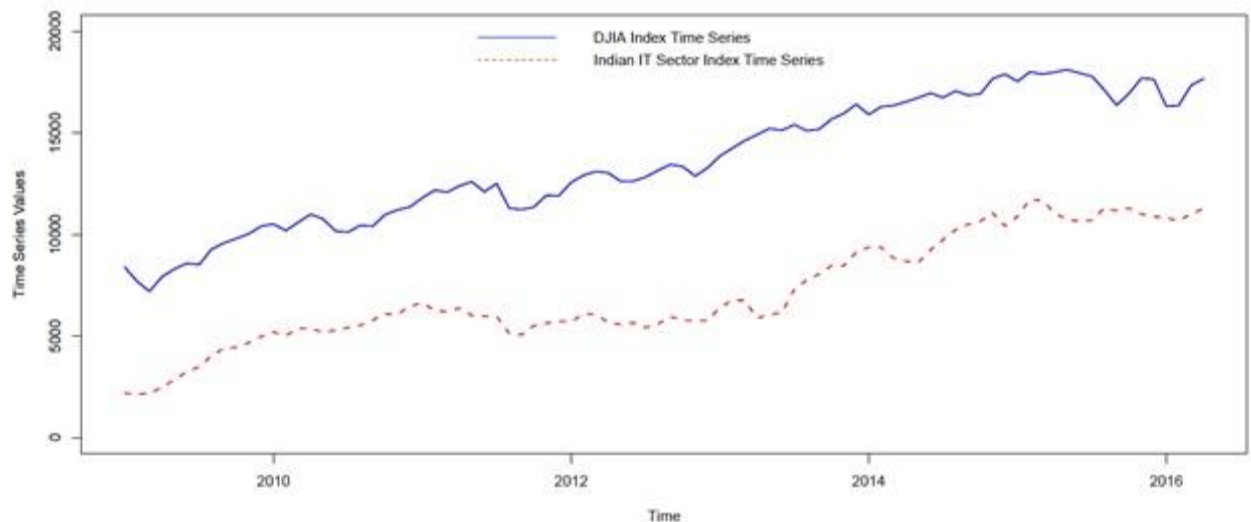

**Figure 11**
Comparison of the aggregate time series of the DJIA index and the Indian IT sector index
(Jan 2009 – Apr 2016)

We have also studied the behavior of the trend components of the two time series during the same period. Figure 12 presents the trend plots of the two time series. It is evident that the trends of the IT time series and the DJIA time series behaved in an identical manner during January 2009 to April 2016. In the similar way, we made a comparative analysis of the behavior of the seasonal components of the two time series. Figure 13 depicts the results obtained. It is clear that the seasonal components of the two time series exhibited similar behavior with the Indian IT seasonality having a lag with respect to the DJIA seasonality. The results in Figures 11, 12 and 13 clearly indicate that Indian IT sector time series has a strong association with the DJIA index time series.

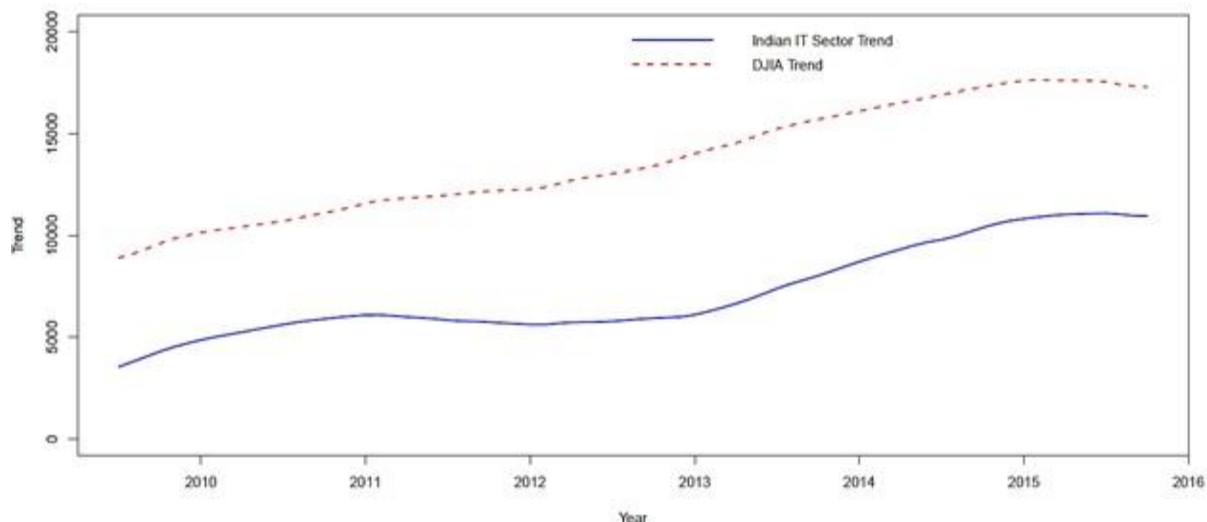

**Figure 12**
Comparison of the trend components of the DJIA index and the Indian IT sector index time series (Jan 2009 – Apr 2016)

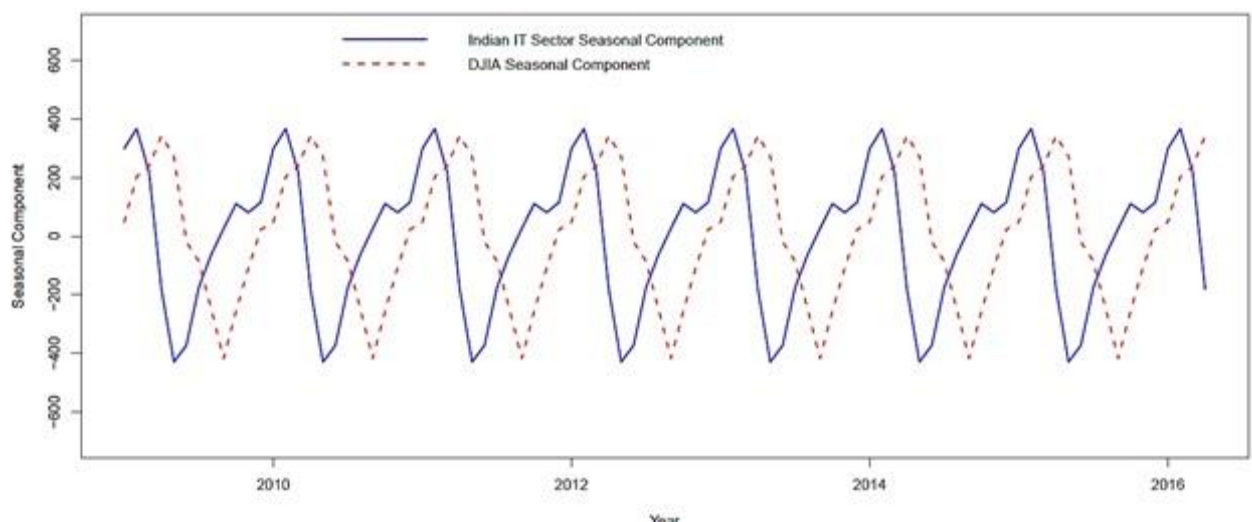

**Figure 13**
Comparison of the seasonal components of the DJIA index and the Indian IT sector index time series (Jan 2009 – Apr 2016)

Having established graphically that the Indian IT sector time series has a strong association with the DJIA time series, we carry out some statistical tests in R in order to prove the association using formal computations. We use *cor.test ( )* function in R to carry out a bivariate correlation test between the IT time series and the DJIA time series. The results of the test are presented in Table 6. The high value (0.945425) of the correlation coefficient with a negligible p-value of the Null hypothesis (of no correlation) implies that the two time series are highly correlated on point-to-point basis.

**Table 6**
Results of the correlation test between Indian IT sector and the DJIA index

| Parameter | Value |
|---|---|
| t- statistic | 26.907 |
| Degrees of freedom (df) | 86 |
| Significance values (p - value) | < 2.2e-16 |
| Correlation coefficient | 0.945425 |

We also study the values of the correlation coefficient at different lags of IT time series with respect to the DJIA time series in order to identify which lag yields the highest value of the correlation coefficient. We used the *ccf ( )* function in R programming environment for this purpose. Figure 14 presents the results. It is evident from Figure 14 that at lag = 0 the highest value of the correlation coefficient is achieved. Hence, it is concluded that the Indian IT time series and the DJIA time series are highly correlated with a zero lag in between them.

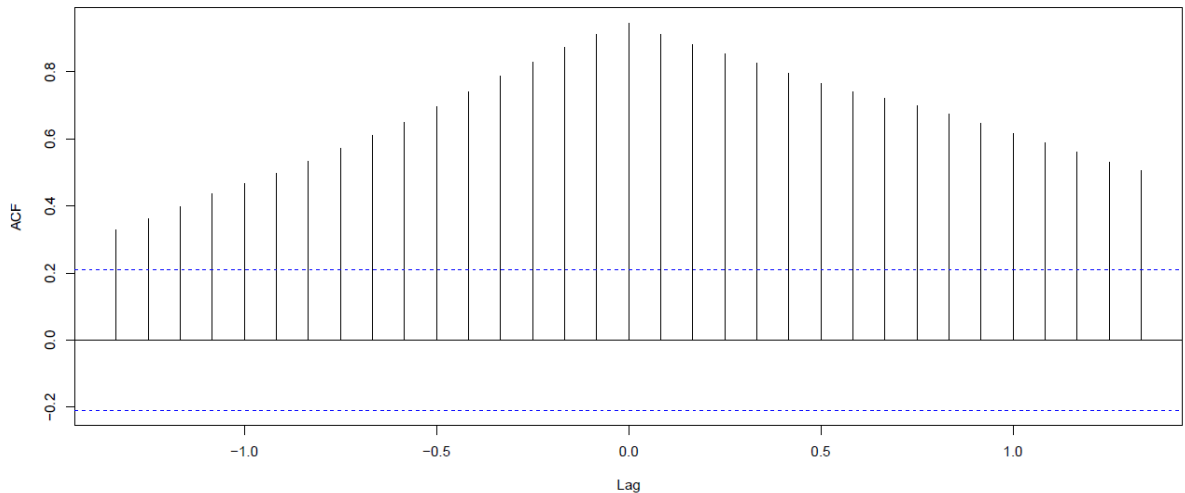

**Figure 14**
The output of the *ccf( )* function depicting the cross correlation between aggregate IT time series and the aggregate DJIA time series (Jan 2009 – Apr 2016)

**Table 7**
Correlation test for the seasonal components of the Indian IT index and the DJIA index

| Parameter | Value |
|---|---|
| t- statistic | -0.47595 |
| Degrees of freedom (df) | 86 |
| Significance values (p - value) | 0.6353 |
| Correlation coefficient | -0.05125503 |

We also studied the association between the seasonal components of the Indian IT time series and the DJIA time series. The results obtained in bivariate correlation test using the *cor.test ( )* function in R environment are presented in Table 7. The extremely small value of the correlation coefficient and the high value of significance indicate that there is no point to point correlation between the seasonal components of the two time series.

In order to identify the lag at which seasonal components of the two time series attain the highest value of the correlation coefficient, we use the *ccf( )* function in R. Figure 15 presents the results. It may be observed that although the correlation is very low (-0.05) at lag value of zero, the cross correlation is approximately around 0.9 (which is quite high) at a lag value of 0.25. Since a lag of 1 represents a time horizon of 12 months, a lag of 0.25 is equivalent to 3 months duration. In other words, the seasonal component of the Indian IT time series has a very strong correlation (e.g., correlation coefficient value approximately 0.9) with the seasonal component of the DJIA time series with a lag of 3 months.

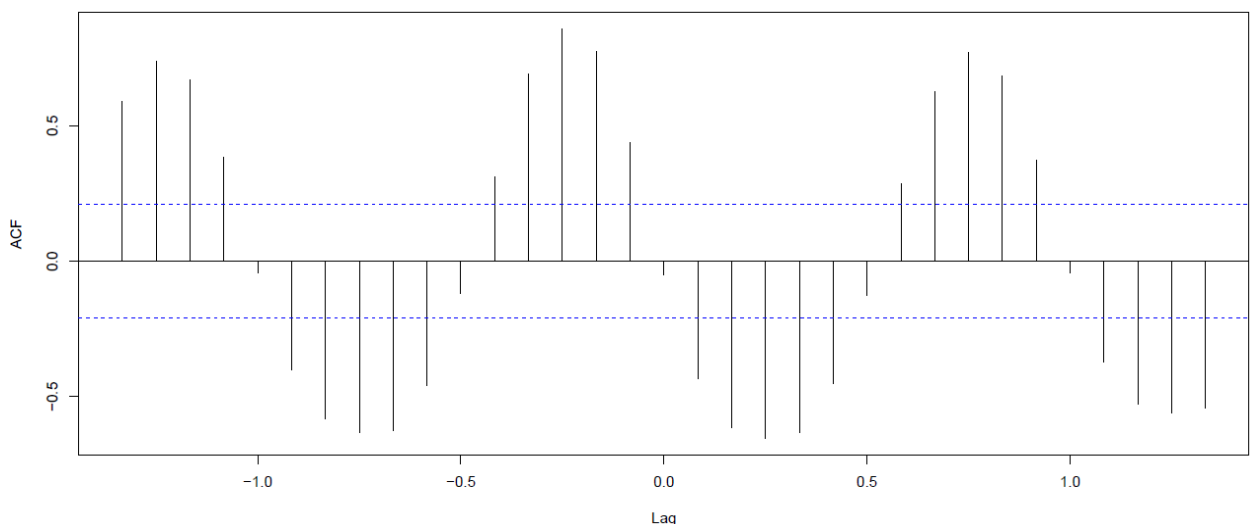

**Figure 15**
The output of the *ccf ()* function depicting the cross correlation between the seasonal components of the IT time series and the DJIA time series

### 4.1.2 Association between Indian IT sector index and Dollar to Rupee exchange rate

In order to study the association between the IT sector time series and the US Dollar to Indian Rupee exchange rate time series, we have first plotted the aggregate time series of both the sectors for the period January 2009 to April 2016. Figure 16 depicts the plot. We have also studied the behavior of the trend components of the two time series of Indian IT sector and the Dollar to Rupee exchange rate. Figure 17 presents the comparison of the trend components. It may be noted that both in Figure 16 and 17, we have multiplied the Dollar to Rupee exchange rate by a factor 0f 100 before plotting in order to make a parity between the ranges of values of the two time series for the purpose of comparison. We do not carry out any study on the seasonality components since for the exchange rate time series, the seasonality does not make any sense.

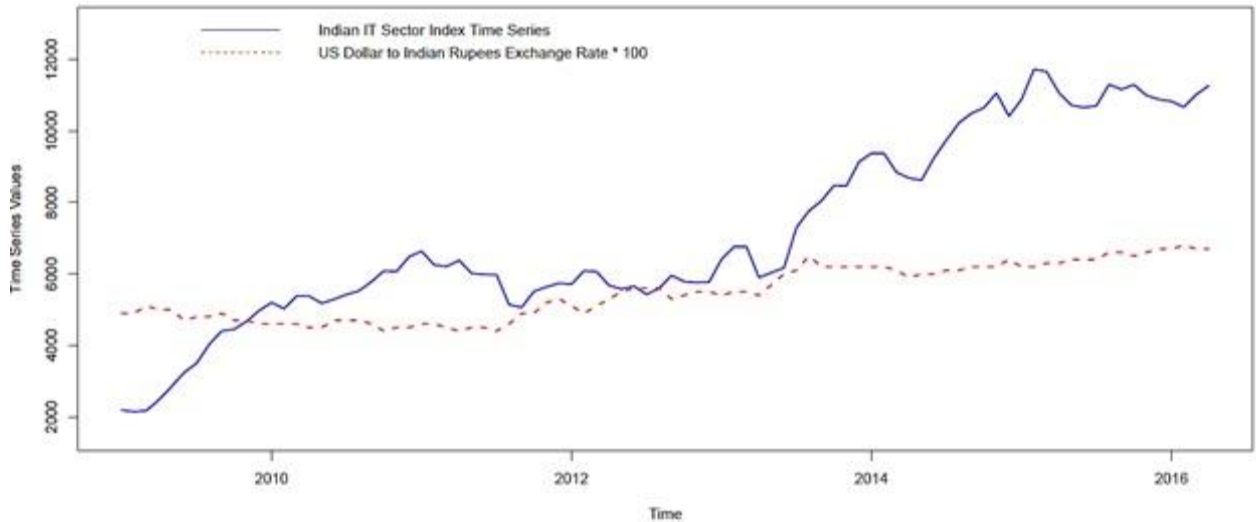

**Figure 16**
Comparison of the aggregate time series of the Indian IT sector index and the US Dollar to Indian Rupees exchange rate (Jan 2009 – Apr 2016)

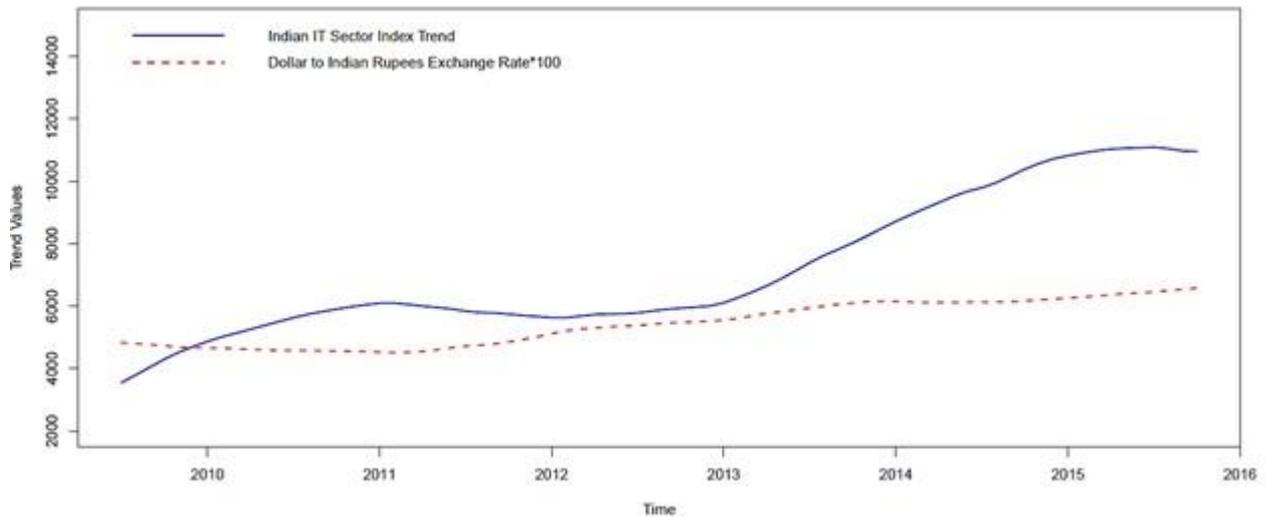

**Figure 17**
Comparison of the trend components of the Indian IT sector index and the US Dollar to Indian Rupees exchange rate time series (Jan 2009 – Apr 2016)

As per our hypotheses, we expect a strong association between the IT time series and the Dollar to Rupee exchange rate time series. A cursory visual inspection of the graphs in Figure 14 and Figure 15 enables us to see a positive association between the two time series. However, we carried out correlation and cross-correlation tests to compute the quantitative values of the association. The results obtained in bivariate correlation test using the *cor.test ( )* function in R environment are presented in Table 8. The high value (0.8333524) of the correlation coefficient with a negligible p-value of the null hypothesis (of no correlation) implies that the two time series are highly correlated on point-to-point basis.

**Table 8**
Correlation test for the Indian IT index and the US Dollar to Indian Rupee exchange rates

| Parameter | Value |
|---|---|
| t- statistic | 13.982 |
| Degrees of freedom (df) | 86 |
| Significance values (p - value) | < 2.2e-16 |
| Correlation coefficient | 0.8333524 |

We also carried out correlation test between the trend components of the IT sector time series and the Dollar to Rupee exchange rate time series. The test yielded a high value (0.8794656) of correlation coefficient with a negligible value of significance level (i.e., the p-value) of less than 2.2 e-15. This clearly indicated a strong point-to-point positive correlation between the two time series and thereby validated our hypothesis that Indian IT sector time series is strongly coupled with the Dollar to Rupee exchange rate time series, both reflecting the world economic picture.

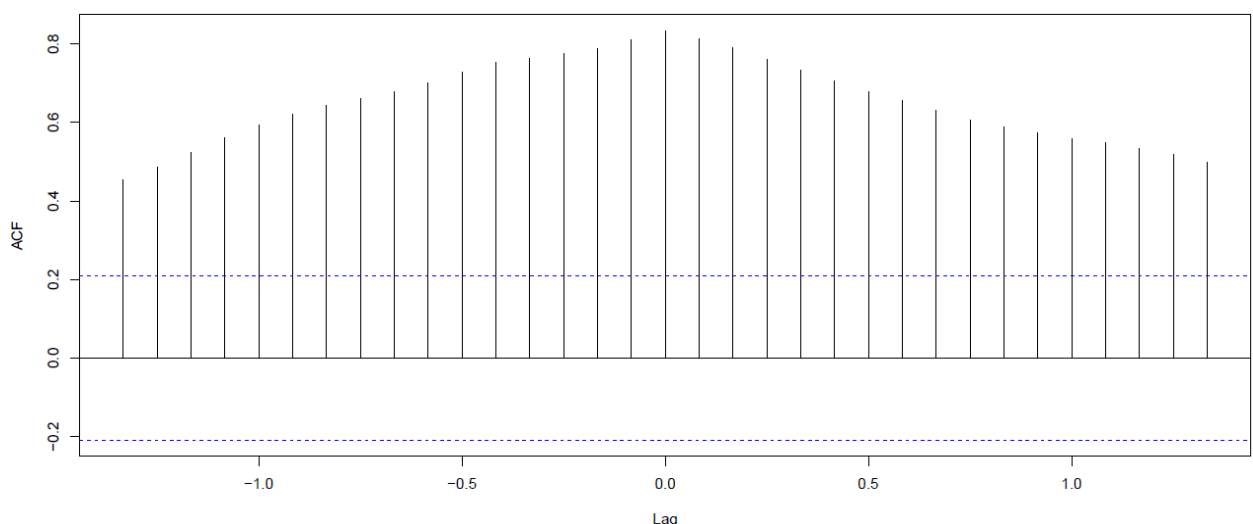

**Figure 18**
The output of the *ccf ( )* function depicting the cross correlation between aggregate IT time series and the aggregate Dollar to Rupee exchange rate

In order to identify the value of the lag that attains the largest magnitude of the correlation coefficient between the IT time series and the Dollar to Rupee exchange rate time series, we used the *ccf ( )* function in R. Figure 16 presents the results. It is evident from Figure 16 that at lag = 0 the highest value of the correlation coefficient is achieved. We also computed the cross-correlation between the trend components of the two time series and also observed that the highest value of the correlation between the time series was obtained at a lag = 0. Hence, it is concluded that the Indian IT time series and the Dollar to Rupee exchange rate time series are highly correlated with a zero lag in between them.

### 4.1.3 Association between Indian IT and NIFTY index time series

The plots of the aggregate time series, the trend components and the seasonal components of the Indian IT sector and the NIFTY index are depicted in Figure 17, Figure 18 and Figure 19 respectively. Even a visual inspection of Figure 17 and Figure 18 gives us an idea that there is a positive association between the IT sector index time series and the NIFTY time series and between their trend components. This validates our hypothesis that the Indian IT sector blue chip stocks have a strong impact on the NIFTY index values, which leads to a positive association between the two index. However, as in the previous cases, we validate our hypothesis by carrying out bivariate correlation tests and cross-correlation tests. Table 9 presents the results of correlation test on the IT sector time series and the NIFTY index time series using the *cor.test ( )* function in R. The high value (0.9609465) of the correlation coefficient with a negligible p-value of the null hypothesis (null hypothesis assumes no correlation) implies that the two time series are highly correlated on point-to-point basis. A correlation test is also carried out between the trend components of the IT sector time series and the NIFTY index time series. The test yielded a even higher value (0.9849783) of correlation coefficient with a negligible value of significance level (i.e., the p-value) of less than 2.2 e-16. This clearly indicated a strong point-to-point positive correlation between the two time series and thereby validated our hypothesis that Indian IT sector time series is strongly coupled with the NIFTY index time series. The cross-correlation study was carried out using the *ccf ( )* function in R. Figure 17 presents the results of the cross-correlation between the IT sector time series and the NIFTY index time series. It is evident that at lag value of zero the highest correlation is attained. This makes it evidently clear that the Indian IT sector and the NIFY index have a very strong point-point positive association between them.

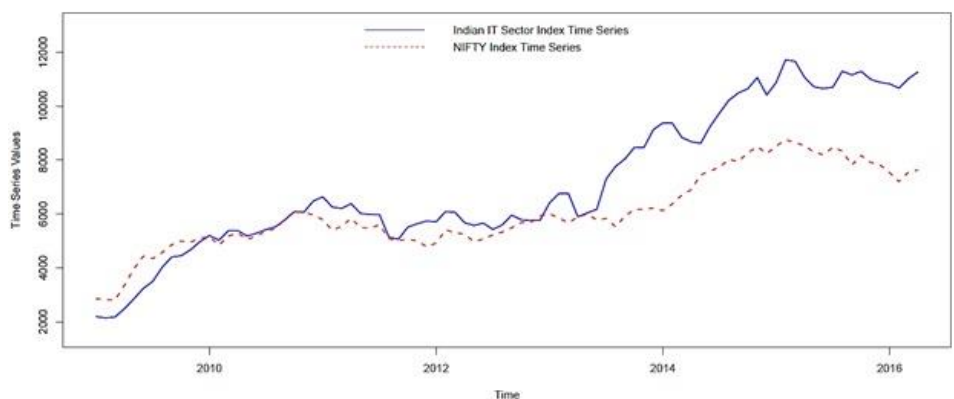

**Figure 19**
Comparison of the aggregate time series of the Indian IT sector index and the NIFTY index
(Jan 2009 – Apr 2016)

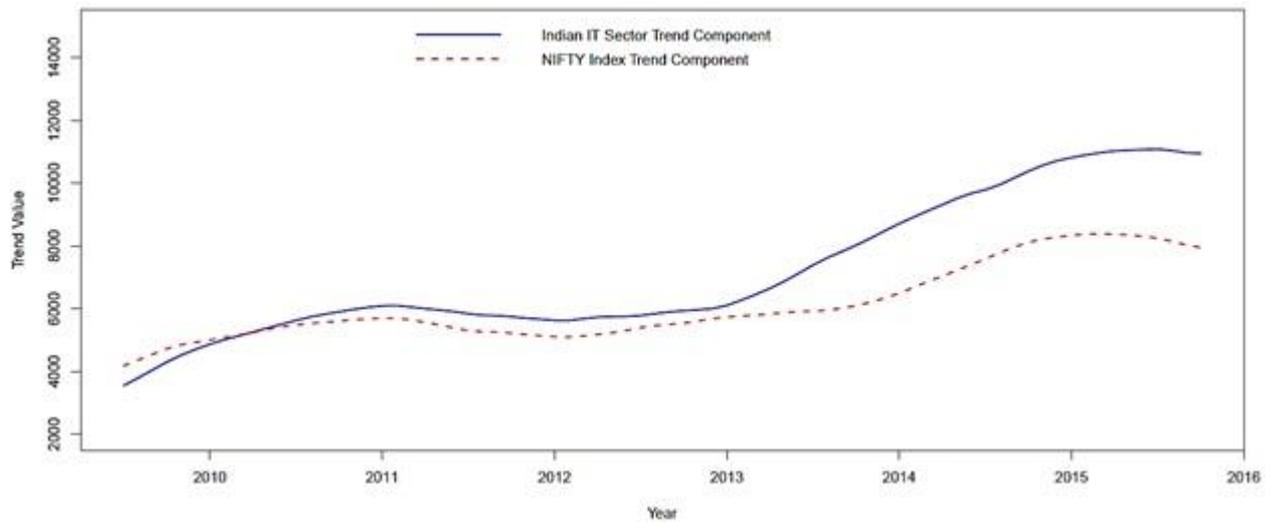

**Figure 20**
Comparison of the trend components of the Indian IT sector and the NIFTY index time series
(Jan 2009 – Apr 2016)

**Table 9**
Correlation test for the Indian IT index and the NIFTY index

| Parameter | Value |
|---|---|
| t- statistic | 32.202 |
| Degrees of freedom (df) | 86 |
| Significance values (p - value) | < 2.2e-16 |
| Correlation coefficient | 0.9609465 |

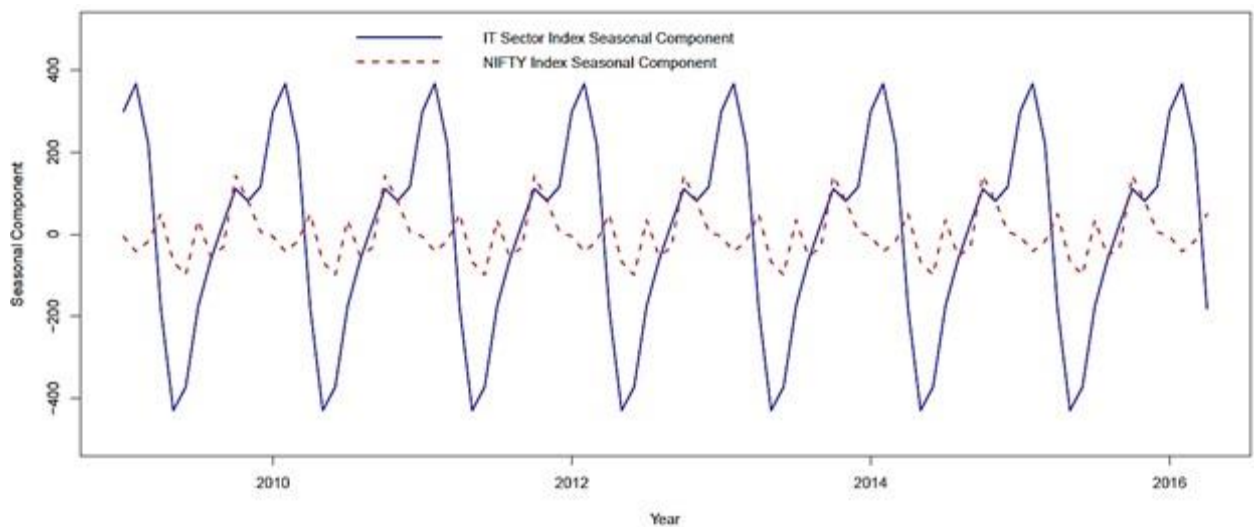

**Figure 21**
Comparison of the seasonal components of the Indian IT sector and the NIFTY index time
series (Jan 2009 – Apr 2016)

In order to investigate the association between the seasonal components of the IT sector time series and the NIFTY index time series, we carry out a correlation and cross correlation study on the seasonal component values of the two time series. Table 10 and Figure 20 present the results of the correlation test and the cross-correlation test.

**Table 10**
Correlation test for the seasonal components of the Indian IT index and the NIFTY index

| Parameter | Value |
|---|---|
| t- statistic | 2.5203 |
| Degrees of freedom (df) | 86 |
| Significance values (p - value) | 0.01357 |
| Correlation coefficient | 0.2622572 |

The very low value of the correlation coeffieicent in Table 10 and low p-value indicate that the sesonal components of the IT sector and the NIFTY index have poor point-to point correlation. Figure 20 depcits the output of the *ccf ( )* function in R and presents the cross-correlation results. It is evident that at lag = 0 the sesonal components have a very low correlation of 0.26. The highest correlation value of approximately 0.47 is achieved at the lag of - 0.4, which is equivalent to 5 months. Both the correlation test and the cross-correlation test results indicate a poor association between the sesonality of the IT sector and the NIFTY index.

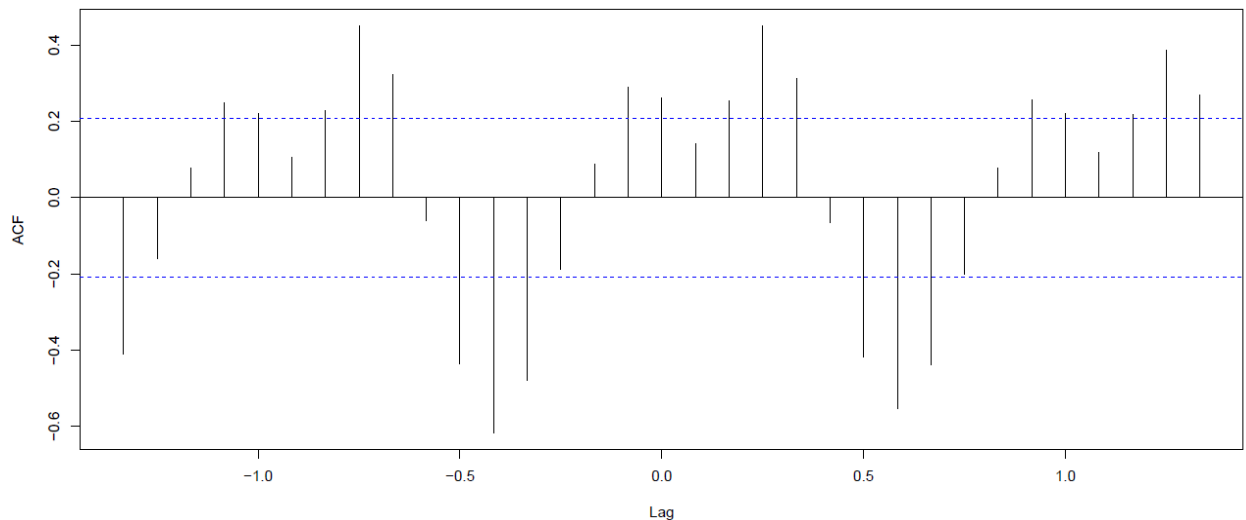

**Figure 22**
The output of the *ccf ( )* function depicting the cross correlation between the seasonal components of the IT sector and the NIFTY time series index

**4.2 Association Analysis of the Indian CG Time Series**

We have tested the association of the Indian CG sector with the DJIA, NIFTY and Dollar to Rupees exchange rate time series in the same way as we have done it for the Indian IT sector in Section 4.1. The detailed results of the study are discussed in this Section.

**4.2.1 Association between the Indian CG and the DJIA time series**

In Figure 21, Figure 22 and Figure 23, we have plotted the Indian CG sector and the DJIA aggregate time series, their trend components and their seasonal components respectively.

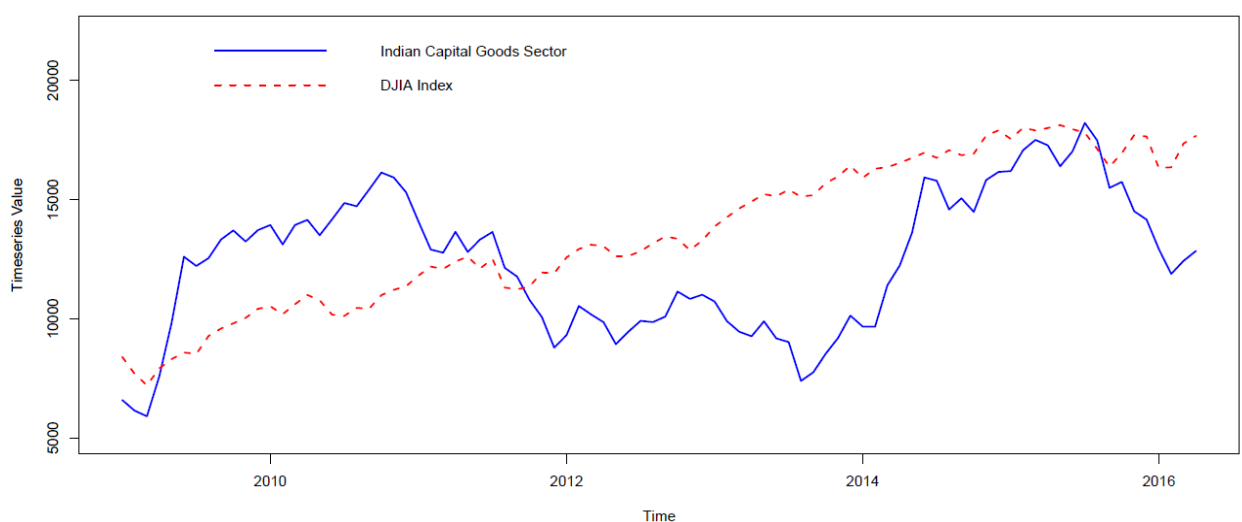

**Figure 23**
Comparison of the aggregate time series of the Indian CG sector index and the DJIA index
(Jan 2009 – 2016)

It is quite evident by visual inspection of Figure 21 and Figure 22 that there is no clear association between the aggregate time series of the two sectors as well as in the time series of their trend components. As in all analyses for the IT sector time series, we carry out the bivariate correlation test and the cross-correlation test for the aggregate time series of the CG sector and the DJIA index. Table 11 and Figure 24 present respectively present the results of the correlation test and the cross-correlation test.

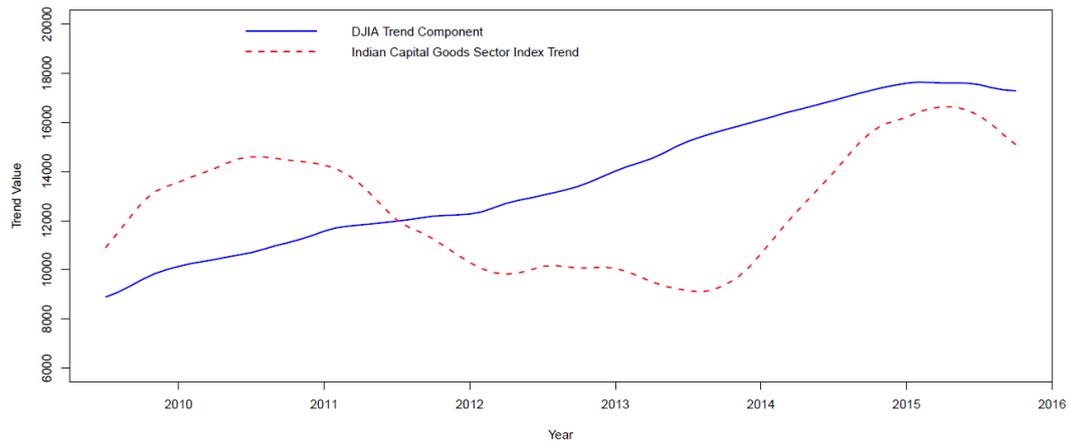

**Figure 24**
Comparison of the trend components of the Indian CG sector index and the DJIA index (Jan 2009 – Apr 2016)

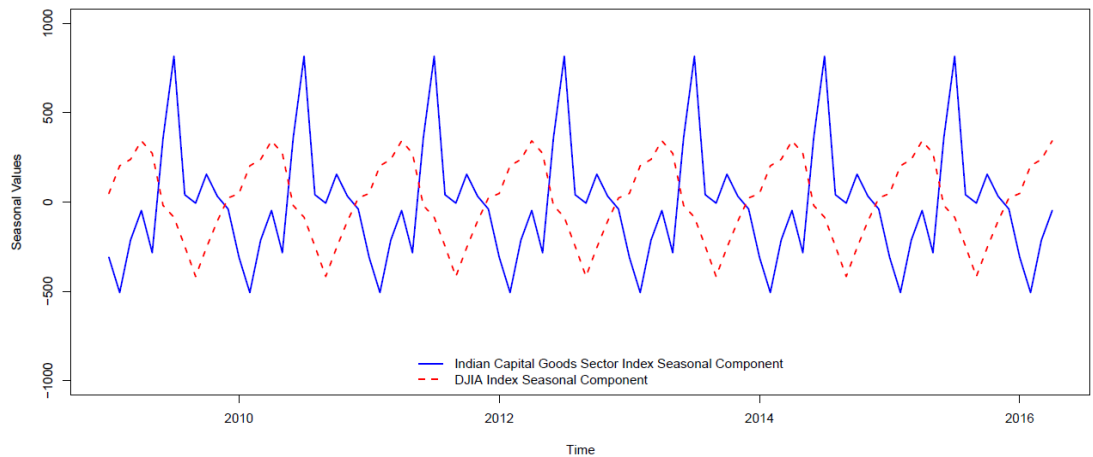

**Figure 25**
Comparison of the seasonal components of the Indian CG sector index and the DJIA index (Jan 2009 – Apr 2016)

**Table 11**
Correlation test for the Indian CG sector index and the DJIA index

| Parameter | Value |
|---|---|
| t- statistic | 0.69046 |
| Degrees of freedom (df) | 70 |
| Significance values (p - value) | 0.4922 |
| Correlation coefficient | 0.08224673 |

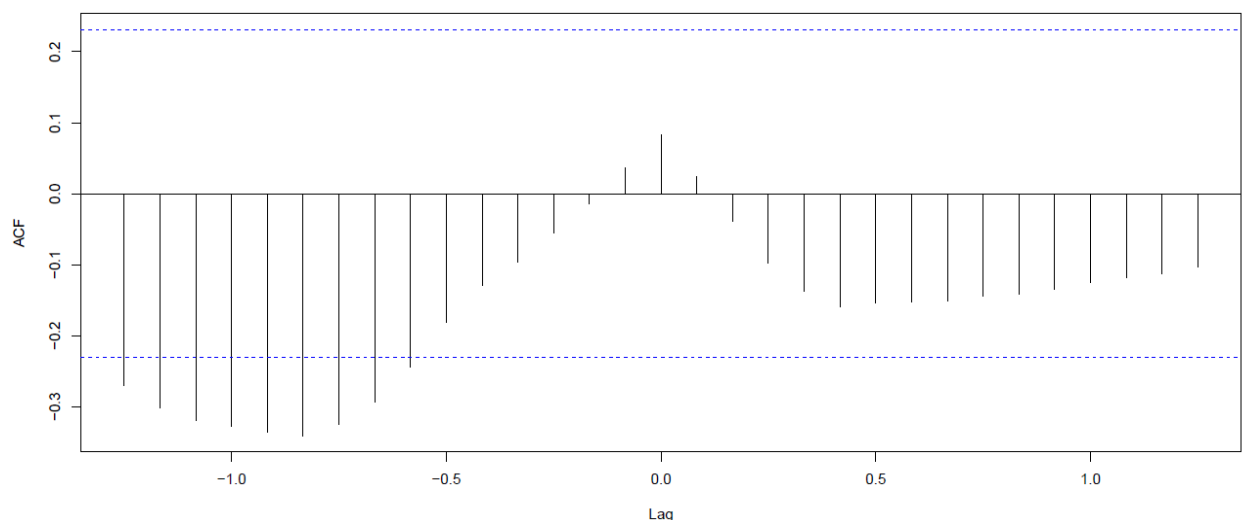

**Figure 26**

The output of the *ccf ( )* function depicting the cross correlation between aggregate CG time series and the aggregate DJIA index time series

    The extremely low value of correlation coefficient (0.08224673) in Table 11 clearly indicates that there is effectively no association between the aggregate time series of the CG sector and the DJIA index. From Figure 24, it is also evident that the numerically largest value of the correlation between the CG and the DJIA aggregate time series is attained at a lag value of nine months. The value of the maximum correlation being -0.32, at a lag of nine months again indicated an absence of any association between the time series. We also carried out the correlation and the cross-correlation tests between the trend components of the two time series. The correlation coefficient between the trend components has been found to be 0.001 with a p-value of 0.05533. The cross-correlation study found that the highest value of correlation was at a lag of 15 months and the value of the correlation coefficient at that lag was approximately 0.27. All these observations indicate that the association between the CG sector time series and the DJIA time series is very poor. This validates our hypothesis that CG sector India is tied to India's domestic growth story and it should have a minimal association with the pattern of growth in DJIA index which essentially reflects the world growth story.

    Since the aggregate time series and the trend components of the Indian CG sector index and the DJIA index were found to have a very poor association, we did not carry out any study on their seasonal components. However, even a very casual look at Figure 23 will make it clear that the seasonal components of the two time series have no association among them.

**4.2.2 Association between Indian CG sector index and Dollar to Rupee exchange rate**

As in all the cases that we have discussed so far, we first plot the behavior of the aggregate time series and their trend components for the Indian CG sector and the US Dollar to Indian Rupee exchange rates. Note that seasonality does not make any sense in the Dollar to Rupee exchange rate time series and hence we do not carry out any seasonality analysis for this comparative study of two time series. Figure 25 and Figure 26 depict respectively the aggregate time series and their trend components for the Indian CG sector index and the Dollar to Rupee exchange rates.

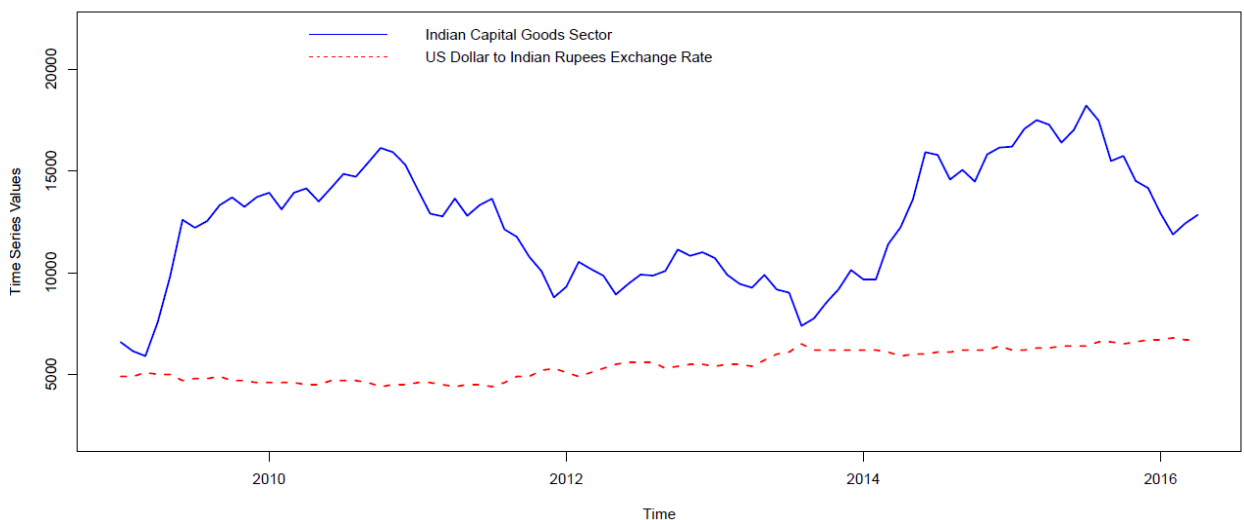

**Figure 27**
Comparison of the aggregate time series of the Indian CG sector index and US Dollar to Indian Rupee exchange rate (Jan 2009 – Apr 2016)

As in the case of comparison of Indian IT sector index time series with the US Dollar to Indian Rupee exchange rate, in both Figure 25 and Figure 26, we have multiplied the Dollar to Rupee exchange rate by a factor of 100 before plotting in order to make a parity between the ranges of values of the two time series for the purpose of comparison. We do not carry out any study on the seasonality components since for the exchange rate time series, the seasonality does not make any sense. Even though a visual inspection of Figure 25 and Figure 26 clearly elicit the fact that there is no perceptible association between the Indian CG sector index time series and the US Dollar to Indian Rupee time series and also between their trends, we carry out bivariate correlation and cross-correlation studies to in order to find the strength of their association. Table 12 and Figure 27 present the results of the correlation and cross-correlation studies respectively.

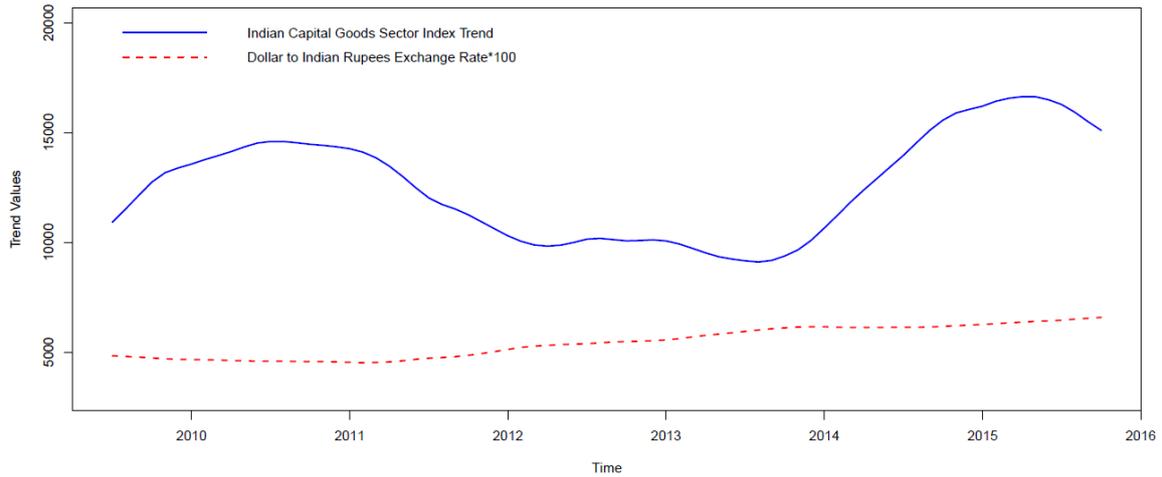

**Figure 28**

Comparison of the trend components of the Indian CG sector index and the US Dollar to Indian Rupee exchange rate (Jan 2009 – Apr 2016)

**Table 12**

Correlation test for the Indian CG index and the US Dollar to Indian Rupee exchange rate

| Parameter | Value |
|---|---|
| t- statistic | 1.028 |
| Degrees of freedom (df) | 86 |
| Significance values (p - value) | 0.3068 |
| Correlation coefficient | 0.1101802 |

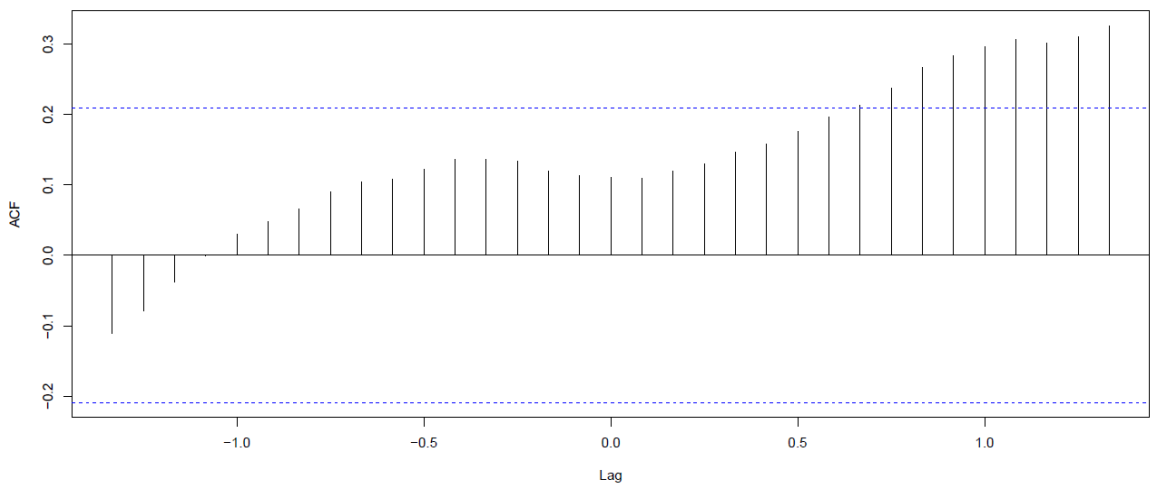

**Figure 29**

The output of the *ccf ( )* function depicting the cross correlation between aggregate CG time series and the aggregate Dollar to Rupee exchange rate time series

From Table 12, we observe that the value of the correlation coefficient (0.11018) is too low with a p-value of 0.3068. Since the p-value is much larger than the threshold value of 0.05, the null hypothesis that assumed no correlation between the time series has got a good support. Hence, the low value of the correlation coefficient and an associated high p-value indicate that the association between the CG sector time series and the US Dollar to Indian Rupee exchange rate is very poor. Figure 27 shows that the highest correlation is attained at a lag of 5 months with the maximum value of the correlation being 0.15. We also carried out correlation and cross-correlation tests on the trend components of the two time series. It has been observed that the trend components of the CG sector and the exchange rate values have a correlation of 0.0951472 with a p-value of 0.4136.The cross-correlation study indicated that the highest correlation was found at a lag of 15 months with the maximum value of the correlation being -0.45. All these observations strongly support our hypothesis that there is no association between the Indian CG sector index time series and the US Dollar to Indian Rupee exchange rate – the CG sector depicting the Indian domestic growth story, while the Dollar to Rupee exchange rate presenting the world economic story.

### 4.2.3   Association between Indian CG sector and NIFTY index time series

We investigate the degree of association between the Indian CG sector index time series and the NIFTY index time series. Our hypothesis is that both the CG sector and the NIFTY relate to the Indian growth story and it is natural that there would a high degree of association between them. We carried out graphical analysis of the two time series by plotting their aggregate index, the trend components and the seasonal components in Figure 28, Figure 29 and Figure 30 respectively.

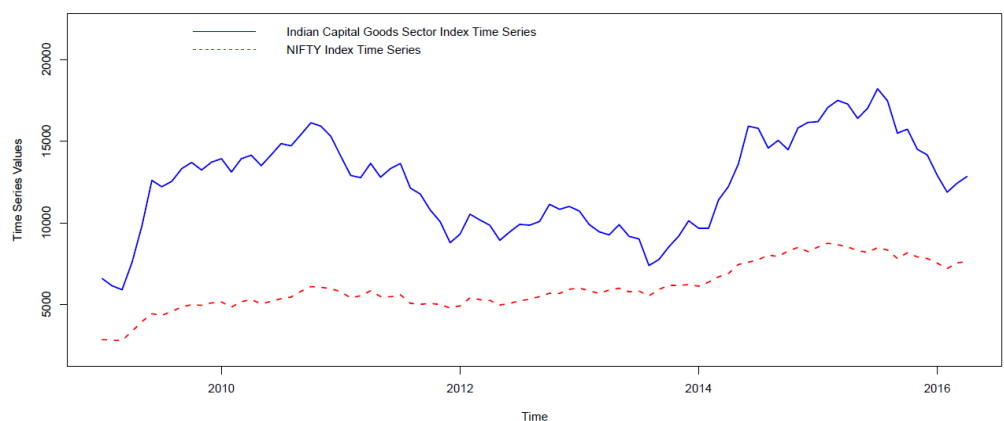

**Figure 30**
Comparison of the aggregate time series of the Indian CG sector index and the NIFTY index
(Jan 2009 – Apr 2016)

Although, from Figure 28 and Figure 29, it possible to get an idea that there is an association between the aggregate time series of the CG sector and the NIFTY index as well as between their trend components, we carried out bivariate correlation and cross-correlation analysis between these time series in order to compute the degree of association between them. Table 13 and Figure 31 present the results of the correlation test and the cross-correlation test respectively.

**Table 13**
Correlation test for the Indian CG sector and the NIFTY index time series

| Parameter | Value |
|---|---|
| t- statistic | 8.8156 |
| Degrees of freedom (df) | 86 |
| Significance values (p - value) | 1.168e-13 |
| Correlation coefficient | 0.6889807 |

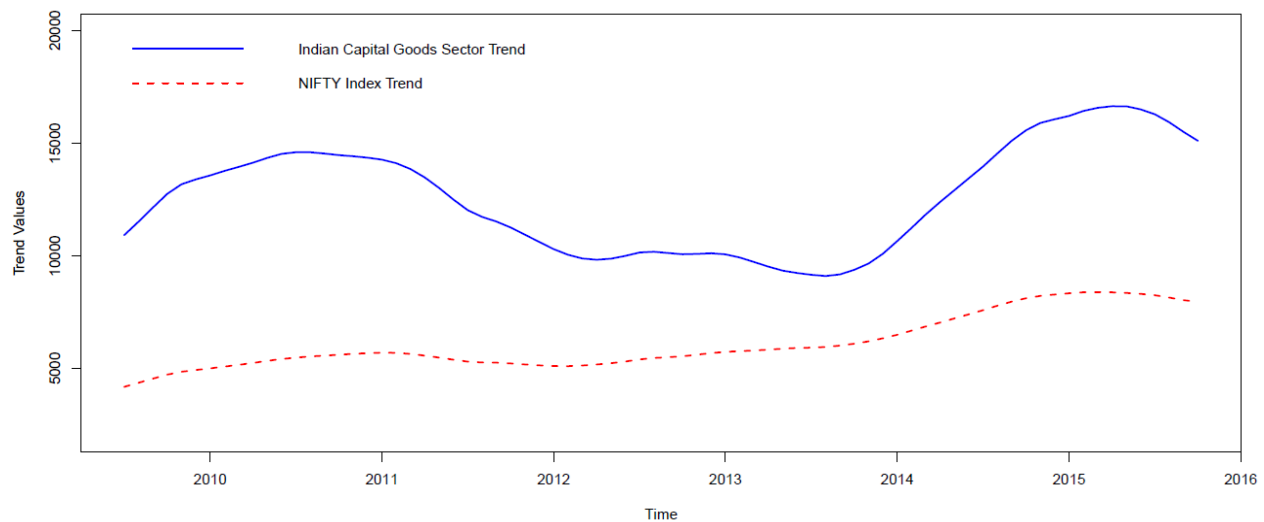

**Figure 31**
Comparison of the trend components of the Indian CG sector and the NIFTY index time series (Jan 2009 – Apr 2016)

From Table 12, it may observed that the value of the correlation coefficient between the overall time series of the CG sector and the overall time series of the NIFTY index is 0.688987 with an associated p-value of which is negligible. Since the p-value indicates the level of support for the null hypothesis that assumes no correlation among the two time series, a reasonably high value of the correlation coefficient and a negligible p-value both indicate a strong association between the two time series. From Figure 31, we observe that the maximum value of correlation coefficient (0.688987) between CG time series and the NIFTY time series is attained at a lag value of zero. This supports the fact that the CG sector time series and the NIFTY index time series has a high point-to-point correlation and hence a strong association between them. We also carried out correlation and cross-correlation tests on the trend components of the two time series. The correlation tests between the trend components yielded a value of correlation coefficient of 0.6052231 with an associated p-value of 6.988e-09. The reasonably high value of the correlation coefficient along with a negligible p-value again indicated a strong association between the trend components of the two time series. The cross-correlation analysis on the trend components showed that the highest value of the correlation (0.6052231) between the trend components of the two time series was at a lag value of zero. All these observations clearly indicated that the CG sector time series and the NIFTY time series have a very strong association between them. This supports our hypothesis that both these sectors essentially depict the domestic growth story of India and hence they should inherently have a strong degree of association among them.

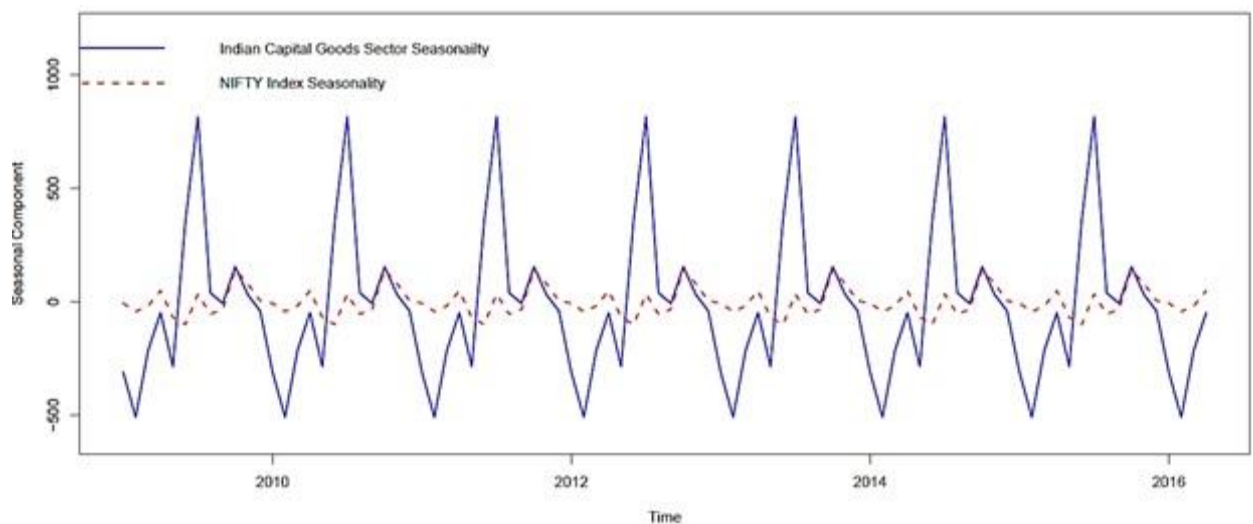

**Figure 32**
Comparison of the seasonal components of the Indian CG sector and the NIFTY index time series (Jan 2009 – Apr 2016)

Since the CG sector time series and the NIFTY index time series and the time series of their trend components are found to have a strong association, we also studied the association between their seasonal components. The visual analysis of Figure 30 being a bit difficult proposition, we carried out correlation and cross-correlation tests on the seasonal components of the two time series. The correlation test yielded a value of 0.2353171 for the correlation coefficient with an associated p-value of 0.02731.These values indicated that the seasonal components of the two time series have a poor point to point correlation. However, as depicted in Figure 32, the cross-correlation study revealed the fact that the CG sector seasonal components has a maximum value of correlation coefficient with the seasonal components of the NIFTY sector at a lag of 4 months with the maximum correlation value being approximately 0.7. This implies that the seasonal components of the two time series have high correlations at certain lags albeit a low point-to-point correlation between them. This further supports our hypothesis of strong association between the Indian CG sector index time series and the NIFTY index time series.

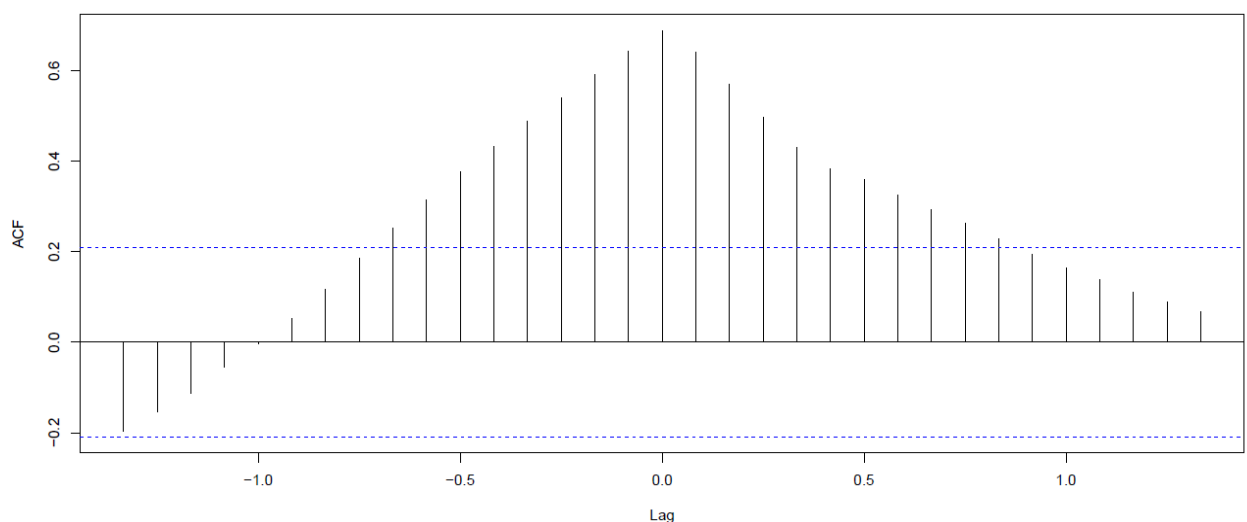

**Figure 33**

The output of the *ccf ( )* function depicting the cross correlation between aggregate CG time series and the aggregate NIFTY index time series

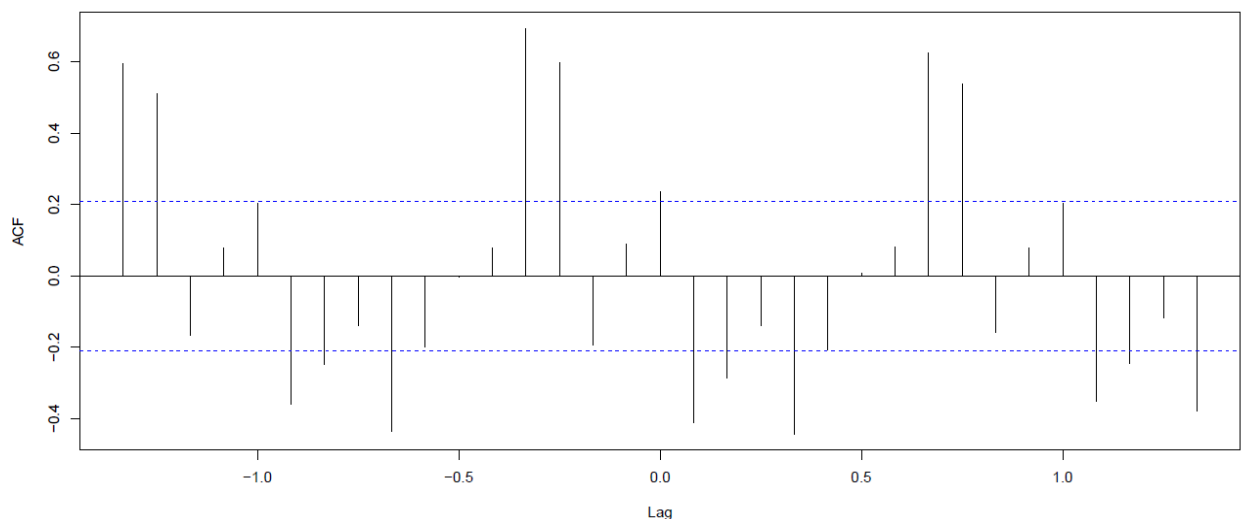

**Figure 34**

The output of the *ccf ( )* function depicting the cross correlation between the seasonal components of the Indian CG sector and the NIFTY time series index

## 5. Forecasting of Time Series Using Linear Regression Model

In Section 4, we have analyzed association among various sectors based on their time series index values. We have found that Indian IT sector index time series has a strong association with the DJIA index, the US Dollar to Indian Rupee exchange rate time series, and also with the NIFTY index time series. While analyzing the Indian CG sector behavior, we observed that the Indian CG sector has a strong association with NIFTY index, but does not have any association with the DJIA index and the US Dollar to Indian Rupee exchange rate. In this Section, we take our analysis to the next level by formulating forecasting models based on linear regression analysis for the time series which exhibited strong association among themselves. We build the following linear regression models for forecasting:

1. Regression of the Indian IT sector index time series on the DJIA index time series and vice-versa.
2. Regression of the Indian IT sector index time series on the US Dollar to Indian Rupee exchange rate time series and vice versa.
3. Regression of the Indian IT sector time series on the NIFTY time series and vice versa.
4. Regression of the Indian CG sector time series on the NIFTY time series and vice versa.

Note that we do not build any model for the Indian CG sector with DJIA index and the CG sector with Dollar to Rupee exchange rate since these pairs of time series were found to have no associations between in them as per our analysis in Section 4.2.1 and Section 4.2.2.

For the purpose of building the linear models between any pair of time series, we suitably divide our datasets into two disjoint sets – one serving as the training data set to construct the model and the other is the test data set to test the effectiveness of the model that is built. For the purpose of training data set construction for any time series, we use the time series data from January 2009 to December 2014. The time series data for the reaming period, i.e., from January 2015 to April 2016 are used for the purpose of test data set. Based on the linear model that is constructed from the training data sets, the values of the dependent time series are forecasted and then compared with the actual values in the test data set. The forecasting error is computed for each month so as to have an idea about the effectiveness of the regression model. In the following sub-sections, we clearly describe the methodology that we have followed in building and testing the forecasting models.

## 5.1  Regression model of Indian IT sector index with the DJIA index time series

In Section 4.1.1, we observed that the bivariate correlation between the Indian IT sector index time series and the DJIA index time series was 0.945425. This indicated that there is a strong linear relationship between the two time series. We now attempt to construct two linear regression models – one for the regression of the Indian IT sector index onto the DJIA index and the other for the regression of the DJIA index onto the Indian IT sector index. While the first model will use the DJIA index as the independent variable and hence will be used for forecasting the Indian IT sector index, the second model will be built using the Indian IT sector index as the independent variable and would be used for forecasting the DJIA index. We use the time series data for both the DJIA index and the Indian IT sector index from January 2009 to December 2014 as the training data set for constructing the model. The time series data from January 2015 to April 2016 are used as the testing data set for evaluating the model efficiency by computing the forecasting error.

For constructing the linear model of the Indian IT sector index on the DJIA index we use the *lm (it_timeseries ~ djia_timeseries)* function call, where the *it_timeseries* parameter stands for the R variable that stores the Indian It sector time series index from January 2009 to December 2014, and the *djia_timeseries* parameter stands for the time series variable that stores the DJIA index for the same period of January 2009 to December 2014. The *lm ( )* function call constructs a linear regression model with the parameter at the left hand side of the *formula* operator ~ being used as the dependent variable, and the parameter at the right hand side of the operator ~ as the independent variable. Using the *summary ( )* function in R, we found some important parameters of the linear regression model of Indian IT index onto the DJIA index. The results are presented in Table 13.

**Table 14**
Summary results of the linear model of the Indian IT index onto the DJIA index

| Parameter | Value |
|---|---|
| Constant term in the regression model | -2585.21 |
| Coefficient of the DJIA index | 0.6958 |
| Residual standard error | 752.4 on 70 degrees of freedom |
| Multiple R- squared | 0.8686 |
| Adjusted R-squared | 0.8667 |

Since the adjusted R-squared value is quite high, we use the linear model to forecast the Indian IT sector index for the period January 2015 to April 2016. We do not bring in any other lag values in the cross-correlation function results since the lag = 0 has a very high point-to-point correlation coefficient between the two time series. We also compare the forecasted IT index with their actual values for each month and compute the forecast error. The results are presented in Table 14. It may be noted that the percentage values of errors are quite nominal.

**Table 15**
Results of forecasting of Indian IT sector index based on DJIA index

| Year | Month | Actual IT Index (A) | Actual DJIA Index (B) | DJIA Index * Coefficient (C) | Constant Term (Intercept) (D) | Forecasted IT Index E = (C + D) | Percent Error (|E − A|/A) *100 |
|---|---|---|---|---|---|---|---|
| 2015 | January | 10882 | 17534 | 12200.16 | -2585.21 | 9615 | 11.64 |
|  | February | 11724 | 18005 | 12527.88 | -2585.21 | 9943 | 15.19 |
|  | March | 11667 | 17891 | 12448.56 | -2585.21 | 9863 | 15.46 |
|  | April | 11073 | 17992 | 12518.83 | -2585.21 | 9934 | 10.29 |
|  | May | 10721 | 18116 | 12605.11 | -2585.21 | 10020 | 6.54 |
|  | June | 10656 | 17945 | 12486.13 | -2585.21 | 9901 | 7.09 |
|  | July | 10703 | 17792 | 12379.67 | -2585.21 | 9794 | 8.49 |
|  | August | 11301 | 17117 | 11910.01 | -2585.21 | 9325 | 17.49 |
|  | September | 11164 | 16367 | 11388.16 | -2585.21 | 8803 | 21.15 |
|  | October | 11296 | 16944 | 11789.64 | -2585.21 | 9204 | 18.52 |
|  | November | 10999 | 17697 | 12313.57 | -2585.21 | 9728 | 11.55 |
|  | December | 10884 | 17639 | 12273.22 | -2585.21 | 9688 | 10.99 |
| 2016 | January | 10832 | 16312 | 11349.89 | -2585.21 | 8765 | 19.06 |
|  | February | 10670 | 16348 | 11374.94 | -2585.21 | 8790 | 17.62 |
|  | March | 11023 | 17344 | 12067.96 | -2585.21 | 9483 | 13.97 |
|  | April | 11270 | 17665 | 12291.31 | -2585.21 | 9706 | 13.88 |

In a similar manner, we constructed a linear model of DJIA index onto the Indian IT sector index for the purpose of forecasting DJIA index based on the Indian IT sector index values. Table 15 presents the output of the *summary ( )* function in R for this model.

Since the adjusted R-squared value is very high, we use this linear model for forecasting the DJIA index for the period January 2015 to April 2016, based on the Indian IT sector index values. The results are presented in Table 16. It may be noted that the magnitudes of the error percentages are quite acceptable.

**Table 16**
Summary results of the linear model of the DJIA index onto the Indian IT index

| Parameter | Value |
|---|---|
| Constant term in the regression model | 4899.25 |
| Coefficient of the IT sector index | 1.248 |
| Residual standard error | 1008 on 70 degrees of freedom |
| Multiple R- squared | 0.8686 |
| Adjusted R-squared | 0.8667 |

**Table 17**
Results of forecasting of the DJIA index based on the Indian IT sector index

| Year | Month | Actual DJIA Index (A) | Actual IT Index (B) | Actual IT Index * Coefficient (C) | Constant Term (Intercept) (D) | Forecasted DJIA Index E = (C + D) | Percent Error (\|E – A\|/A) *100 |
|---|---|---|---|---|---|---|---|
| 2015 | January | 17534 | 10882 | 12200.16 | 4899.25 | 18480 | 5.40 |
| | February | 18005 | 11724 | 12527.88 | 4899.25 | 19531 | 8.47 |
| | March | 17891 | 11667 | 12448.56 | 4899.25 | 19460 | 8.77 |
| | April | 17992 | 11073 | 12518.83 | 4899.25 | 18718 | 4.04 |
| | May | 18116 | 10721 | 12605.11 | 4899.25 | 18279 | 0.90 |
| | June | 17945 | 10656 | 12486.13 | 4899.25 | 18198 | 1.41 |
| | July | 17792 | 10703 | 12379.67 | 4899.25 | 18257 | 2.62 |
| | August | 17117 | 11301 | 11910.01 | 4899.25 | 19003 | 11.02 |
| | September | 16367 | 11164 | 11388.16 | 4899.25 | 18832 | 15.06 |
| | October | 16944 | 11296 | 11789.64 | 4899.25 | 18997 | 12.11 |
| | November | 17697 | 10999 | 12313.57 | 4899.25 | 18626 | 5.25 |
| | December | 17639 | 10884 | 12273.22 | 4899.25 | 18482 | 4.78 |
| 2016 | January | 16312 | 10832 | 11349.89 | 4899.25 | 18418 | 12.91 |
| | February | 16348 | 10670 | 11374.94 | 4899.25 | 18215 | 11.42 |
| | March | 17344 | 11023 | 12067.96 | 4899.25 | 18656 | 7.56 |
| | April | 17665 | 11270 | 12291.31 | 4899.25 | 18964 | 7.35 |

**5.2 Regression model of the Indian IT sector index with Dollar to Rupee exchange rate**

In Section 4.1.2, we observed that the Indian IT sector index time series had a strong positive association with the US Dollar to Indian Rupee exchange rate time series. The correlation coefficient between the two time series was found to be 0.8333524. Hence, we build two linear models for these two time series – one for the regression of Indian IT sector index onto the Dollar to Rupee exchange rate, and the other for the regression of Dollar to Rupee exchange rate onto the Indian IT sector index. Since the methodology for building the models remains identical to the one discussed in Section 5.1, we present only the results of the two models.

Table 17 presents the output of the *summary ( )* function for the linear model of the Indian IT sector index time series onto the Dollar to Rupee exchange rate time series.

**Table 18**
Summary results of the linear model of the Indian IT sector index onto the US Dollar to Indian Rupee exchange rate

| Parameter | Value |
|---|---|
| Constant term in the regression model | -5722.8 |
| Coefficient of the Dollar to Rupee exchange rate | 228.3 |
| Residual standard error | 1451 on 70 degrees of freedom |
| Multiple R- squared | 0.5113 |
| Adjusted R-squared | 0.5045 |

**Table 19**
Results of forecasting of Indian IT sector index based on US Dollar to Indian Rupee exchange rates

| Year | Month | Actual IT Index (A) | Actual Exchange Rate (B) | Actual Exch Rate * Coefficient (C) | Constant Term (Intercept) (D) | Forecasted IT Index E = (C + D) | Percent Error (\|E – A\|/A) *100 |
|---|---|---|---|---|---|---|---|
| 2015 | January | 10882 | 62 | 14154.6 | -5722.8 | 8432 | 22.52 |
| | February | 11724 | 62 | 14154.6 | -5722.8 | 8432 | 28.08 |
| | March | 11667 | 63 | 14382.9 | -5722.8 | 8660 | 25.77 |
| | April | 11073 | 63 | 14382.9 | -5722.8 | 8660 | 21.79 |
| | May | 10721 | 64 | 14611.2 | -5722.8 | 8888 | 17.09 |
| | June | 10656 | 64 | 14611.2 | -5722.8 | 8888 | 16.59 |
| | July | 10703 | 64 | 14611.2 | -5722.8 | 8888 | 16.95 |
| | August | 11301 | 66 | 15067.8 | -5722.8 | 9345 | 17.30 |
| | September | 11164 | 66 | 15067.8 | -5722.8 | 9345 | 16.29 |
| | October | 11296 | 65 | 14839.5 | -5722.8 | 9116 | 19.29 |
| | November | 10999 | 66 | 15067.8 | -5722.8 | 9345 | 15.04 |
| | December | 10884 | 67 | 15296.1 | -5722.8 | 9573 | 12.04 |
| 2016 | January | 10832 | 67 | 15296.1 | -5722.8 | 9573 | 11.62 |
| | February | 10670 | 68 | 15524.4 | -5722.8 | 9801 | 8.14 |
| | March | 11023 | 67 | 15296.1 | -5722.8 | 9573 | 13.15 |
| | April | 11270 | 67 | 15296.1 | -5722.8 | 9573 | 15.06 |

The results of forecasting for the Indian IT sector index for the period January 2015 to April 2016 based on the US Dollar to Indian Rupee exchange rate have been presented in Table 18. The errors in forecasting are found to be quite nominal considering the fact that exchange rate time series has a small range of variations compared to the variations in the Indian IT sector index value. We also constructed a linear model of the US Dollar to Indian Rupee exchange rate time series on to the Indian IT sector index time series. Table 19 presents the output of the *summary ( )* function for the linear model. This model has been used to forecast the Dollar to Rupee exchange rates based on the Indian IT sector index values.

**Table 20**

Summary results of the linear model of the US Dollar to Indian Rupee exchange rate onto the Indian IT sector index

| Parameter | Value |
|---|---|
| Constant term in the regression model | 38.50 |
| Coefficient of the IT sector index | 0.002240 |
| Residual standard error | 4.546 on 70 degrees of freedom |
| Multiple R- squared | 0.5113 |
| Adjusted R-squared | 0.5043 |

**Table 21**

Results of forecasting of the US Dollar to Indian Rupee exchange rate based on Indian IT sector index

| Year | Month | Actual Exchange Rate (A) | Actual IT Index (B) | Actual IT Index * Coefficient (C) | Constant Term (Intercept) (D) | Forecasted Exch. Rate E = (C + D) | Percent Error (\|E − A\|/A) *100 |
|---|---|---|---|---|---|---|---|
| 2015 | January | 62 | 10882 | 24.38 | 38. 50 | 63 | 1.41 |
| | February | 62 | 11724 | 26.26 | 38. 50 | 65 | 4.45 |
| | March | 63 | 11667 | 26.13 | 38. 50 | 65 | 2.59 |
| | April | 63 | 11073 | 24.80 | 38. 50 | 63 | 0.00 |
| | May | 64 | 10721 | 24.02 | 38. 50 | 63 | 2.33 |
| | June | 64 | 10656 | 23.87 | 38. 50 | 62 | 2.39 |
| | July | 64 | 10703 | 23.97 | 38. 50 | 62 | 2.39 |
| | August | 66 | 11301 | 25.31 | 38. 50 | 64 | 3.32 |
| | September | 66 | 11164 | 25.01 | 38. 50 | 64 | 3.32 |
| | October | 65 | 11296 | 25.30 | 38. 50 | 64 | 1.85 |
| | November | 66 | 10999 | 24.64 | 38. 50 | 63 | 4.34 |
| | December | 67 | 10884 | 24.38 | 38. 50 | 63 | 6.16 |
| 2016 | January | 67 | 10832 | 24.26 | 38. 50 | 63 | 6.16 |
| | February | 68 | 10670 | 23.90 | 38. 50 | 62 | 8.24 |
| | March | 67 | 11023 | 24.69 | 38. 50 | 63 | 5.69 |
| | April | 67 | 11270 | 25.24 | 38. 50 | 64 | 4.87 |

Since the adjusted R-squared value is 0.5043, we investigated whether there was any possibility to increase it by bringing in values at some other lags. However, we found that by bringing in time series values at other lags, we could not improve the adjusted R-squared value. Hence we used the model in Table 19 for forecasting the US Dollar to Indian Rupee exchange rates for the period January 2015 to April 2016 based on the Indian IT sector index during the same period. The results are presented in Table 20. The percentage values of the forecast errors are found to be quite low.

*J. Sen et al. / Journal of Insurance and Financial Management, Vol. 1, Issue 4 (2016) 68-131*    115## 5.3 Regression model of Indian IT sector index with NIFTY index

In Section 4.1.3, we observed that the Indian IT sector index had a very strong positive association with the NIFTY index. The correlation coefficient between the two time series was found to be 0.9609465. Therefore, we build to linear models for these two time series.

Table 21 presents the output of the *summary ( )* function of the linear model of the Indian IT sector index onto the NIFTY index. We will use this model to forecast IT sector index for the period January 2015 to April 2016 based on the NIFTY index values during the same period.

**Table 22**
Summary results of the linear model of the Indian IT index onto the NIFTY index

| Parameter | Value |
|---|---|
| Constant term in the regression model | -3179.63 |
| Coefficient of the NIFTY index | 1.685 |
| Residual standard error | 709.6 on 70 degrees of freedom |
| Multiple R- squared | 0.8831 |
| Adjusted R-squared | 0.8815 |

**Table 23**
Results of forecasting of the Indian IT sector index based on the NIFTY index

| Year | Month | Actual IT Index (A) | NIFTY Index (B) | NIFTY Index * Coefficient (C) | Constant Term (Intercept) (D) | Forecasted IT Index E = (C + D) | Percent Error (\|E − A\|/A) *100 |
|---|---|---|---|---|---|---|---|
| 2015 | January | 10882 | 8518 | 14352.83 | -3179.63 | 11173 | 2.68 |
|  | February | 11724 | 8750 | 14743.75 | -3179.63 | 11564 | 1.36 |
|  | March | 11667 | 8664 | 14598.84 | -3179.63 | 11419 | 2.12 |
|  | April | 11073 | 8524 | 14362.94 | -3179.63 | 11183 | 1.00 |
|  | May | 10721 | 8300 | 13985.50 | -3179.63 | 10806 | 0.79 |
|  | June | 10656 | 8196 | 13810.26 | -3179.63 | 10631 | 0.24 |
|  | July | 10703 | 8477 | 14283.75 | -3179.63 | 11104 | 3.75 |
|  | August | 11301 | 8337 | 14047.85 | -3179.63 | 10868 | 3.83 |
|  | September | 11164 | 7816 | 13169.96 | -3179.63 | 9990 | 10.51 |
|  | October | 11296 | 8169 | 13764.77 | -3179.63 | 10585 | 6.29 |
|  | November | 10999 | 7913 | 13333.41 | -3179.63 | 10154 | 7.68 |
|  | December | 10884 | 7818 | 13173.33 | -3179.63 | 9994 | 8.18 |
| 2016 | January | 10832 | 7536 | 12698.16 | -3179.63 | 9519 | 12.13 |
|  | February | 10670 | 7200 | 12132.00 | -3179.63 | 8952 | 16.10 |
|  | March | 11023 | 7550 | 12721.75 | -3179.63 | 9542 | 13.43 |
|  | April | 11270 | 7632 | 12859.92 | -3179.63 | 9680 | 14.11 |

Since the adjusted R-squared value in Table 21 is quite high, we used the linear model in Table 21 without investigating any other lags in the cross-correlation function. The linear model is used for forecasting Indian IT sector index for the period January 2015 to April 2016 based on the NIFTY index during that period. The forecasting results are presented in Table 22. It is evidently clear that the percentage values of the forecast errors are quite small.

We also built a linear model of the NIFTY index onto the Indian IT sector index time series. Table 23 depicts the output of the *summary ( )* function of the model. We have used this model to forecast the NIFTY index for the period January 2015 to April 2016 based on the Indian IT sector index during the same period.

**Table 24**
Summary results of the linear model of the NIFTY index onto the Indian IT index

| Parameter | Value |
|---|---|
| Constant term in the regression model | 2321.47 |
| Coefficient of the IT sector index | 0.524 |
| Residual standard error | 395.7 on 70 degrees of freedom |
| Multiple R- squared | 0.8831 |
| Adjusted R-squared | 0.8815 |

**Table 25**
Results of forecasting of the NIFTY index based on the Indian IT sector index

| Year | Month | Actual NIFTY Index (A) | Actual IT Index (B) | Actual IT Index * Coefficient (C) | Constant Term (Intercept) (D) | Forecasted NIFTY Index E = (C + D) | Percent Error (\|E − A\|/A) *100 |
|---|---|---|---|---|---|---|---|
| 2015 | January | 8518 | 10882 | 5702.17 | 2321.47 | 8024 | 5.80 |
| | February | 8750 | 11724 | 6143.38 | 2321.47 | 8465 | 3.26 |
| | March | 8664 | 11667 | 6113.51 | 2321.47 | 8435 | 2.64 |
| | April | 8524 | 11073 | 5802.25 | 2321.47 | 8124 | 4.70 |
| | May | 8300 | 10721 | 5617.80 | 2321.47 | 7939 | 4.35 |
| | June | 8196 | 10656 | 5583.74 | 2321.47 | 7905 | 3.55 |
| | July | 8477 | 10703 | 5608.37 | 2321.47 | 7930 | 6.45 |
| | August | 8337 | 11301 | 5921.72 | 2321.47 | 8243 | 1.13 |
| | September | 7816 | 11164 | 5849.94 | 2321.47 | 8171 | 4.55 |
| | October | 8169 | 11296 | 5919.10 | 2321.47 | 8241 | 0.88 |
| | November | 7913 | 10999 | 5763.48 | 2321.47 | 8085 | 2.17 |
| | December | 7818 | 10884 | 5703.22 | 2321.47 | 8025 | 2.64 |
| 2016 | January | 7536 | 10832 | 5675.97 | 2321.47 | 7997 | 6.12 |
| | February | 7200 | 10670 | 5591.08 | 2321.47 | 7913 | 9.90 |
| | March | 7550 | 11023 | 5776.05 | 2321.47 | 8098 | 7.26 |
| | April | 7632 | 11270 | 5905.48 | 2321.47 | 8227 | 7.80 |

Since the linear model has yielded an adjusted R-squared value that is quite high, we do not consider any other lag in the cross-correlation function between the two time series and use the linear model in Table 23 to forecast the NIFTY index based on the Indian IT sector index. The results of forecasting are presented in Table 24. The percentage values of the forecast errors are found be quite low and hence acceptable.

**5.4 Regression model of the Indian CG sector index with the NIFTY index**

In Section 4.2.3, we observed that the Indian CG sector index have a fair degree of association with the NIFTY index. The correlation coefficient between the two time series over the period January 2009 to April 2016 was found to be 0.6889807. Based on this fair degree of association between the two time series, we have built to linear models for the purpose of forecasting the values of one time series given the other and vice versa.

**Table 26**
Summary results of the linear model of the Indian CG index onto the NIFTY index
(The training data set for the model is based on data for the period Jan 2009 – Dec 2014)

| Parameter | Value |
|---|---|
| Constant term in the regression model | 4756.20 |
| Coefficient of the NIFTY index | 1.2444 |
| Residual standard error | 2251 on 70 degrees of freedom |
| Multiple R- squared | 0.2905 |
| Adjusted R-squared | 0.2804 |

**Table 27**
Results of forecasting of Indian CG sector index based on the NIFTY index (The model used data for the period Jan 2009 – Dec 2014 as the training data set)

| Year | Month | Actual CG Index (A) | Actual NIFTY Index (B) | NIFTY Index * Coefficient (C) | Constant Term (Intercept) (D) | Forecasted CG Index E = (C + D) | Percent Error (\|E – A\|/A) *100 |
|---|---|---|---|---|---|---|---|
| 2015 | January | 16189 | 8518 | 10596.39 | 4756.20 | 15353 | 5.16 |
| | February | 17064 | 8750 | 10885.00 | 4756.20 | 15641 | 8.34 |
| | March | 17495 | 8664 | 10778.02 | 4756.20 | 15534 | 11.21 |
| | April | 17266 | 8524 | 10603.86 | 4756.20 | 15360 | 11.04 |
| | May | 16389 | 8300 | 10325.20 | 4756.20 | 15081 | 7.98 |
| | June | 17013 | 8196 | 10195.82 | 4756.20 | 14952 | 12.11 |
| | July | 18206 | 8477 | 10545.39 | 4756.20 | 15302 | 15.95 |
| | August | 17471 | 8337 | 10371.23 | 4756.20 | 15127 | 13.42 |
| | September | 15485 | 7816 | 9723.10 | 4756.20 | 14479 | 6.50 |
| | October | 15734 | 8169 | 10162.24 | 4756.20 | 14918 | 5.19 |
| | November | 14501 | 7913 | 9843.77 | 4756.20 | 14600 | 0.68 |
| | December | 14155 | 7818 | 9725.59 | 4756.20 | 14482 | 2.31 |
| 2016 | January | 12914 | 7536 | 9374.78 | 4756.20 | 14131 | 9.42 |
| | February | 11875 | 7200 | 8956.80 | 4756.20 | 13713 | 15.48 |
| | March | 12424 | 7550 | 9392.20 | 4756.20 | 14148 | 13.88 |
| | April | 12837 | 7632 | 9494.21 | 4756.20 | 14250 | 11.01 |

Table 25 depicts the output of the *summary ( )* function of the linear model of the Indian CG sector index onto the NIFTY index. We have used the time series data of both the time series for the period January 2009 to December 2014 for constructing the training data set of the model. Since the value of the adjusted R-squared in Table 25 is quite low, we investigated whether it could be improved upon by considering the time series values at other lags in the cross-correlation function. However, we failed in our attempt as the value of the adjusted R-squared could not be increased. Hence, we used the linear model in Table 25 to forecast the Indian CG sector index for the period January 2015 to April 2016 based on the NIFTY index over the same period. Table 26 presents the forecasting results. It is quite clear that the percentage values of forecast errors are low and within the acceptable range.

With the objective of forecasting the NIFTY index based on the Indian CG sector time series index, we build a linear model of the NIFTY index onto the Indian CG sector index. The output of the *summary ( )* of this linear model is depicted in Table 27. Note that the training data set for the model is based on both the time series data for the period January 2009 to December 2014. We used the linear model in Table 27 to forecast the NIFTY index for the period January 2015 to April 2016 based on the Indian CG sector index over the same period.

**Table 28**
Summary results of the linear model of the NIFTY index onto the Indian CG index
(The training data set for the model is based on data for the period Jan 2009 – Dec 2014)

| Parameter | Value |
| --- | --- |
| Constant term in the regression model | 2867.72 |
| Coefficient of the Indian CG sector index | 0.2335 |
| Residual standard error | 974.9 on 70 degrees of freedom |
| Multiple R- squared | 0.2905 |
| Adjusted R-squared | 0.2804 |

Table 28 represents the forecasting results of the NIFTY index based on the India CG sector index for the period January 2015 to April 2016. The percentage values of the forecasting errors are found to be a bit high although within the acceptable range of less than 25%.

Although the percentage values of forecasting errors in both Table 26 and Table 28 are within acceptable limits, we explored ways to improve upon the forecasting accuracies by suitably changing the model construction methods. A careful look at Figure 2 indicates that the CG index time series has undergone a number of changes in its pattern over the period January 2009 to April 2016. Similarly, the NIFTY index time series also has undergone a number of changes in its behavior over the same period of time as evident from Figure 4. Linear models constructed from these changing time series over a long time horizon of 6 years (January 2009 to December 2014) will inherently be far from perfect. We observe that both the Indian CG sector index and the NIFTY index time series had experienced consistent upward trends during the entire year of 2014 and both have exhibited somewhat downward swings in the year of 2015 onwards. We utilized this observation in choosing our training data set for constructing the linear models between the CG sector index and the NIFTY index. For constructing the linear models, we leave out the time series index of both the sectors from January 2009 till December 2013 and use the data of 2014 as the training data set. The objective is to build the models based on the most recent observations leaving out past data over a long time horizon.

**Table 29**
Results of forecasting of the NIFTY index based on Indian CG sector index (The model used data for the period Jan 2009 – Dec 2014 as the training data set)

| Year | Month | Actual NIFTY Index (A) | Actual CG Index (B) | Actual CG Index * Coefficient (C) | Constant Term (Intercept) (D) | Forecasted NIFTY Index E = (C + D) | Percent Error (\|E – A\|/A) *100 |
|---|---|---|---|---|---|---|---|
| 2015 | January | 8518 | 16189 | 3780.13 | 2867.72 | 6648 | 21.95 |
| | February | 8750 | 17064 | 3984.44 | 2867.72 | 6852 | 21.69 |
| | March | 8664 | 17495 | 4085.08 | 2867.72 | 6953 | 19.75 |
| | April | 8524 | 17266 | 4031.61 | 2867.72 | 6899 | 19.06 |
| | May | 8300 | 16389 | 3826.83 | 2867.72 | 6695 | 19.34 |
| | June | 8196 | 17013 | 3972.54 | 2867.72 | 6840 | 16.54 |
| | July | 8477 | 18206 | 4251.10 | 2867.72 | 7119 | 16.02 |
| | August | 8337 | 17471 | 4079.48 | 2867.72 | 6947 | 16.67 |
| | September | 7816 | 15485 | 3615.75 | 2867.72 | 6483 | 17.05 |
| | October | 8169 | 15734 | 3673.89 | 2867.72 | 6542 | 19.92 |
| | November | 7913 | 14501 | 3385.98 | 2867.72 | 6254 | 20.97 |
| | December | 7818 | 14155 | 3305.19 | 2867.72 | 6173 | 21.04 |
| 2016 | January | 7536 | 12914 | 3015.42 | 2867.72 | 5883 | 21.93 |
| | February | 7200 | 11875 | 2772.81 | 2867.72 | 5641 | 21.65 |
| | March | 7550 | 12424 | 2901.00 | 2867.72 | 5769 | 23.59 |
| | April | 7632 | 12837 | 2997.44 | 2867.72 | 5865 | 23.15 |

Table 29 presents the output of the *summary ( )* function of the newly constructed linear model of the Indian CG sector onto the NIFTY index. Note that the model is constructed using the time series index of the Indian CG sector and the NIFTY from January 2014 to December 2014 as the training data set. It is interesting to observe that the adjusted R-squared value has increased to 0.8322 from 0.2804 (from Table 25) just by changing the training data set to include only the most recent observations. The linear model in Table 29 is used for forecasting the Indian CG sector index for the period January 2015 to April 2016 based on the NIFTY index for the same period. Table 30 presents the results of forecasting. A visual inspection of the error values in Table 26 and Table 30 makes it evidently clear that the forecast accuracy has improved substantially in the new model.

**Table 30**
Summary results of the linear model of the Indian CG index onto the NIFTY index
(The training data set for the model is based on data for the period Jan 2014 – Dec 2014)

| Parameter | Value |
|---|---|
| Constant term in the regression model | -7174.22 |
| Coefficient of the NIFTY index | 2.7870 |
| Residual standard error | 979.8 on 10 degrees of freedom |
| Multiple R- squared | 0.8474 |
| Adjusted R-squared | 0.8322 |

**Table 31**
Results of forecasting of the Indian CG sector index based on the NIFTY index
(The model used data for the period Jan 2014 – Dec 2014 as the training data set)

| Year | Month | Actual CG Index (A) | Actual NIFTY Index (B) | NIFTY Index * Coefficient (C) | Constant Term (Intercept) (D) | Forecasted CG Index E = (C + D) | Percent Error (\|E – A\|/A) *100 |
|---|---|---|---|---|---|---|---|
| 2015 | January | 16189 | 8518 | 23739.67 | -7174.22 | 16565 | 2.33 |
| | February | 17064 | 8750 | 24386.25 | -7174.22 | 17212 | 0.87 |
| | March | 17495 | 8664 | 24146.57 | -7174.22 | 16972 | 2.99 |
| | April | 17266 | 8524 | 23756.39 | -7174.22 | 16582 | 3.96 |
| | May | 16389 | 8300 | 23132.10 | -7174.22 | 15958 | 2.63 |
| | June | 17013 | 8196 | 22842.25 | -7174.22 | 15668 | 7.91 |
| | July | 18206 | 8477 | 23625.40 | -7174.22 | 16451 | 9.64 |
| | August | 17471 | 8337 | 23235.22 | -7174.22 | 16061 | 8.07 |
| | September | 15485 | 7816 | 21783.19 | -7174.22 | 14609 | 5.66 |
| | October | 15734 | 8169 | 22767.00 | -7174.22 | 15593 | 0.90 |
| | November | 14501 | 7913 | 22053.53 | -7174.22 | 14879 | 2.61 |
| | December | 14155 | 7818 | 21788.77 | -7174.22 | 14615 | 3.25 |
| 2016 | January | 12914 | 7536 | 21002.83 | -7174.22 | 13829 | 7.08 |
| | February | 11875 | 7200 | 20066.40 | -7174.22 | 12892 | 8.57 |
| | March | 12424 | 7550 | 21041.85 | -7174.22 | 13868 | 11.62 |
| | April | 12837 | 7632 | 21270.38 | -7174.22 | 14096 | 9.81 |

Table 31 presents the output of the *summary ( )* function for the new linear model of the NIFTY index onto the Indian CG sector index. The model is constructed using the time series index of the NIFTY index and the Indian CG sector index from January 2014 to December 2014 as the training data set. We observe that the adjusted R-squared value has increased to 0.8322 from 0.2804 (from Table 27) because of the inclusion of only the most recent observations of the two time series in the training data set of the model.

**Table 32**
Summary results of the linear model of the NIFTY index onto the Indian CG index
(The training data set for the model is based on data for the period Jan 2014 – Dec 2014)

| Parameter | Value |
|---|---|
| Constant term in the regression model | 3323.79 |
| Coefficient of the Indian CG sector index | 0.3041 |
| Residual standard error | 323.6 on 10 degrees of freedom |
| Multiple R- squared | 0.8474 |
| Adjusted R-squared | 0.8322 |

The linear model in Table 31 is used for forecasting the NIFTY index for the period January 2015 to April 2016 based on the Indian CG sector index for the same period. Table 32 presents the results of forecasting. A visual inspection of the error values in Table 28 and Table 32 makes it evidently clear that the forecast accuracy has improved substantially in the new model.

**Table 33**
Results of forecasting of the NIFTY index based on Indian CG sector index
(The model used data for the period Jan 2014 – Dec 2014 as the training data set)

| Year | Month | Actual NIFTY Index (A) | Actual CG Index (B) | Actual CG Index * Coefficient (C) | Constant Term (Intercept) (D) | Forecasted NIFTY Index E = (C + D) | Percent Error (\|E − A\|/A) *100 |
|---|---|---|---|---|---|---|---|
| 2015 | January | 8518 | 16189 | 4923.08 | 3323.79 | 8246.87 | 3.18 |
|  | February | 8750 | 17064 | 5189.16 | 3323.79 | 8512.95 | 2.71 |
|  | March | 8664 | 17495 | 5320.23 | 3323.79 | 8644.02 | 0.23 |
|  | April | 8524 | 17266 | 5250.59 | 3323.79 | 8574.38 | 0.59 |
|  | May | 8300 | 16389 | 4983.90 | 3323.79 | 8307.69 | 0.09 |
|  | June | 8196 | 17013 | 5173.65 | 3323.79 | 8497.44 | 3.68 |
|  | July | 8477 | 18206 | 5536.45 | 3323.79 | 8860.24 | 4.52 |
|  | August | 8337 | 17471 | 5312.93 | 3323.79 | 8636.72 | 3.60 |
|  | September | 7816 | 15485 | 4708.99 | 3323.79 | 8032.78 | 2.77 |
|  | October | 8169 | 15734 | 4784.71 | 3323.79 | 8108.50 | 0.74 |
|  | November | 7913 | 14501 | 4409.75 | 3323.79 | 7733.54 | 2.27 |
|  | December | 7818 | 14155 | 4304.54 | 3323.79 | 7628.33 | 2.43 |
| 2016 | January | 7536 | 12914 | 3927.15 | 3323.79 | 7250.94 | 3.78 |
|  | February | 7200 | 11875 | 3611.19 | 3323.79 | 6934.98 | 3.68 |
|  | March | 7550 | 12424 | 3778.14 | 3323.79 | 7101.93 | 5.93 |
|  | April | 7632 | 12837 | 3903.73 | 3323.79 | 7227.52 | 5.30 |

## 6. Related Work

Researchers have spent considerable effort in designing mechanisms for forecasting of daily stock prices. Applications of neural network based approaches have been proposed in many forecasting systems. The main advantage of neural networks is that they can approximate any nonlinear function to an arbitrary degree of accuracy with a suitable number of hidden units (Hornik et al., 1989).

Zhang et al. propose the application of a well-known neural network technique - multilayer back propagation (BP) neural network - in financial data mining (Zhang et al., 2004). The authors present a modified neural network-based forecasting model and an intelligent mining system. The system can forecast the buying and selling signs according to the prediction of future trends of stock market, and provide decision-making for stock investors. The simulation results of seven years of Shanghai composite index show that the return achieved by the system is about three times higher than that achieved by the *buy-and-hold* strategy.

Accurate volatility forecasting is the core task in risk management in which various portfolios' pricing, hedging, and option strategies are exercised. Roh proposes a hybrid model with neural network and various time series models for forecasting the volatility of stock price index from two viewpoints: deviation and direction (Roh, 2007). The results demonstrate the utility of the hybrid model for volatility forecasting.

Mostafa proposed neural network-based mechanism to predict stock market movements in Kuwait using data for the period January 2001 to December 2003 (Mostafa, 2010). The author used two neural network architectures: *multi-layer perception* (MLP) neural networks and generalized regression neural networks to predict the Kuwait Stock Exchange (KSE) closing price movements. It had been demonstrated that due to their robustness and flexibility of modeling algorithms, neuro-computational models usually outperform traditional statistical techniques such as regression and ARIMA in forecasting price movements in stock exchanges.

Kimoto et al applied neural networks on historical accounting data and used various macroeconomic parameters for the purpose of prediction of variations in stock returns (Kimoto et al, 1990). The authors presented a system for buying and selling timing prediction system for stocks in the Tokyo Stock Exchange. The learning algorithms proposed by the authors were demonstrated to produce accurate predictions of stock prices.

Leigh et al proposed the use of linear regression and simple neural network models for forecasting the stock market index in the New York Stock Exchange during the period 1981-1999 (Leigh et al, 2005). The authors proposed a mechanism to identify trading volume spikes through the use of a template matching technique based on statistical pattern recognition. The days that matched the condition signifying volume spikes, application of linear regression was done to model the future change in price using historical price and prime interest rate values.

Hammad et al. demonstrated that *artificial neural network* (ANN) model can be trained to converge to an optimal solution while it maintains a very high level of precision in forecasting of stock prices (Hammad et al., 2009). The authors developed a back propagation algorithm for forecasting Jordanian stock prices using feed forward multi-layer neural network. MATLAB simulations conducted on seven Jordanian companies from service and manufacturing sectors produced extremely accurate predictions of stock prices.

Dutta et al demonstrate the application of ANN models for forecasting Bombay Stock Exchange's SENSEX weekly closing values for the period of January 2002 to December 2003 (Dutta et al, 2006). The authors developed two networks with three hidden layers. The first network was provided with the inputs of the weekly closing value, 52-week moving average of the weekly closing SENSEX values, 5-week moving average of the closing values, and the 10-week oscillator for the past 200 weeks. The second network was supplied with the inputs of weekly closing value, 52-week moving average of the weekly closing SENSEX values, 5-week moving average of the closing values and the 5-week volatility for the past 200 weeks. The performance of the two networks were evaluated by measuring their root mean square error values and the mean absolute error values in prediction of the weekly closing SENSEX values for the period January 2002 to December 2003.

Tsai and Wang proposed a mechanism of combining ANN and decision tree-based approach to create a stock price forecasting model (Tsai & Wang, 2009). The experimental results demonstrated that the combined ANN and decision tree-based approach had higher forecast accuracy than the single ANN and the single decision tree-based approach.

Tseng et al. deployed traditional time series decomposition (TSD), HoltWinters (H/W) models, Box-Jenkins (B/J) methodology and neural network- based approach on 50 randomly chosen stocks during September 1, 1998 - December 31, 2010 resulting in a total of 3105 observations for each company's closing stock prices (Tseng et al, 2012). For hold-out period or out-of-sample forecasts over 60 trading days, it has been observed that the forecasting errors are lower for B/J, H/W and normalized neural network model, while the errors are appreciably larger for time series decomposition and non-normalized neural network models.

Moshiri and Cameron presented a Back Propagation Network (BPN) with econometric models to forecast inflation using (i) Box-Jenkins Autoregressive Integrated Moving Average (ARIMA) model, (ii) Vector Autoregressive (VAR) model and (iii) Bayesian Vector Autoregressive (BVAR) model (Moshiri, & Cameron, 2010). The authors compared each of these three approaches with a hybrid BPN model using the same set of variables. Forecasts were made over three different time horizons: one, three and twelve months. The performances of the models were compared using their root mean squared errors and mean absolute errors. It had been found that the hybrid BPN models outperformed the other approaches in many cases.

Phua et al deployed ANNs with genetic algorithms for the purpose of predicting the stock prices in Singapore Stock Exchange (Phua et al, 2000). The result was promising with a forecast accuracy of 81% on the average.

Hutichinson et al proposed a learning network-based non-parametric method for estimating the pricing formula of a derivative (Hutchinson et al, 1994). Some of the primary economic variables that influence the derivative price, e.g., the current fundamental asset price, the strike price, the time to maturity etc. were used as the inputs to the learning network. The derivative price was determined by the output into which the learning network mapped the inputs. The daily closing prices of S&P 500 futures and the options for the period from January 1987 to December 1991 were used as the training data set to construct the model. For the purpose of comparing the relative performance of various models, the authors used four methods: ordinary least squares, radial basis function networks, multilayer perceptron networks, and the projection pursuit. The non-parametric model was found to be more accurate in making derivative pricing forecasts than its parametric pricing counterpart.

Chen et al. have proposed an approach for constructing a model for predicting the direction of return on the Taiwan Stock Exchange Index (Chen et al., 2003). The authors argued that trading strategies guided by forecasts of the direction of price movement are more effective and usually lead to higher profit. Probabilistic Neural Network (PNN) was used to forecast the direction of index return after it was trained using historical data. The authors applied their forecasts to various index trading strategies, of which the performances were compared with those generated by the buy and hold strategy, and the investment strategies guided by the forecasts estimated by the random walk model and the parametric Generalized Method of Moments (GMMM) with Kalman filter.

Basalto et al. applied a pair-wise clustering approach to the analysis of the Dow Jones Index companies, in order to identify similar temporal behavior of the traded stock prices (Basalto et al., 2005). The objective of the authors was to explore the dynamics which that govern the companies' stock prices. A pairwise version of the chaotic map algorithm had been deployed that used the correlation coefficients between the financial time series to find the similarity measures for clustering the temporal patterns. The coupling interactions between the maps are considered as the functions the correlation coefficients. The resultant dynamics of such systems formed the clusters of companies that belong to different industrial branches. These clusters of companies can be gainfully exploited to optimize portfolio construction.

Chen et al. studied how the seemingly chaotic behavior of stock markets could be very well represented using Local Linear Wavelet Neural Network (LLWNN) technique (Chen et al, 2005). The LLWNN was optimized by using Estimation of Distribution Algorithm (EDA). The objective was to predict the share price for the following trade day based on the opening, closing and maximum values of the stock price on a given day. The results showed that even in seemingly random fluctuations, there was an underlying deterministic feature that was directly enciphered in the opening, closing and the maximum values of the index of any day thereby making predictability quite possible.

de Faria et al. proposed predictive framework for the principal index of the Brazilian stock market through ANN and adaptive exponential smoothing method (de Faria et al., 2009). The objective of the study was to compare the forecasting accuracies of both methods on the Brazilian stock index, with a particular focus on prediction of the sign of the market returns. Results showed that both methods were equally efficient in predicting the index returns. However, the neural networks outperformed the adaptive exponential smoothing method in forecasting of the market movement, with relative hit rates similar to ones found in other developed markets.

Hanias et al. conducted a study to predict the daily stock exchange price index of the Athens Stock Exchange (ASE) using a neural network with back propagation (Hanias et al., 2012). Multistep prediction for nine days ahead was achieved using the network with a very low value of 0.0024 for the minimum Mean Square Error (MSE). This also indicated that the same approach could be gainfully applied for daily and weekly prediction of stock prices.

Ning et al. proposed a chaotic neural network-based scheme for stock index prediction (Ning et al., 2009). The author used data from a Chinese stock market and a Shenzhen stock market. The chaotic neural network was used to learn the non-linear stochastic and chaotic patterns in the stock market indices and based on its learning the network was used to forecast future indices of the stock market.

Shen et al. presented a tapped delay neural network (TDNN) with an adaptive learning and pruning algorithm for predicting non-linear time series of stock index (Shen et al., 2007). The authors trained the TDNN by recursive least square (RLS) in which the learning-rate parameter could be chosen automatically so that the network convergence was achieved very fast. The resultant trained network was optimized by utilizing a pruning algorithm leading to reduction in computational complexity. The simulation results showed that the optimized network not only reduced the computational complexity but also achieved improved prediction accuracies.

Zhu et al. carried out a study to reinforce the hypothesis that there is a significant bidirectional nonlinear causality between stock returns and trading volumes (Zhu et al., 2008). In their approach, the authors used a component-based neural network in forecasting one-step ahead stock index increments. The model was further enriched by inclusion of different combinations of index and component stocks' trading volumes as inputs. The network was trained using data of stock returns and volumes from NASDAQ, DJIA and STI index. The experimental results demonstrated that the augmented neural networks with trading volumes lead to improvements in forecasting performance under different terms of forecasting horizon.

Wang & Nie proposed a combination of three forecasting techniques such as the grey model (GM (1, 1)), BP neural networks and support vector machines (SVM) to forecast the Shanghai Industrial Index, the Shanghai Commercial Index, the Shanghai Real Estate Index, and the Shanghai Public Utilities Index (Wang & Nie, 2008). The authors then used optimal weight linear combination of forecasts model, the BP neural-based combination forecast model and the SVM-based combination forecasts model for the purpose of carrying out forecasts on the same index. The results indicated that combination forecasts model could improve the forecast accuracies to a large extent.

Wu et al. proposed an ensemble model of SVM and ANNs for the purpose of predicting three stock index (Wu et al., 2008). The authors have compared the forecasting performance of the ensemble model with those of the SVM model and the ANN model. The results clearly indicated that the performance of the ensemble approach was superior to those of the other two models.

Bentes et al. have investigated the long memory and volatility clustering for the S&P 500, NASDAQ 100 and Stoxx 50 indexes in order to compare the US and European markets (Bentes et al., 2008). The main objective of the authors were to compare two different perspectives: the traditional approach in which the authors have considered the GARCH (1, 1), IGARCH(1, 1) and FIGARCH (1, d, 1) specifications and the econophysics approach based on the concept of entropy. The authors had chosen three variants of this notion: the Shannon, Renyi and Tsallis measures. The results from both perspectives had shown nonlinear dynamics in the volatility of SP 500, NASDAQ 100 and Stoxx 50 indexes.

Liao et al. investigated stock market investment issues on Taiwan stock market using a two-stage data mining approach (Liao et al., 2008). In the first stage, the authors have used the *apriori* algorithm to mine association rules and knowledge patterns about stock category association and possible stock category investment collections. After mining of association rules and knowledge patterns, the K-means clustering algorithm were used to identify the stock clusters in order to make a robust categorization of stocks in the Taiwan stock market. Several possible stock market portfolio alternatives under different situations had also been proposed.

A number of work have been carried out by researchers using time series and fuzzy time series approaches for handling forecasting problems. Thenmozhi examined the nonlinear nature of the Bombay Stock Exchange time series using chaos theory (Thenmozhi, 2001). The study examined the Sensex returns time series from August 1980 to September 1997 and showed that the daily returns and weekly returns of the BSE sensex are characterized by nonlinearity and the time series is weakly chaotic.

Fu et al. have proposed a method of representing a financial time series according to the importance of data points (Fu et al., 2007). Using the concepts of data point importance, the authors have constructed a tree data structure that supports incremental updating of the time series. The technique enables the user to present a time series in different levels of details and also facilitates multi-resolution dimensionality reduction of a large time series data. The authors have presented several evaluation methods of data point importance, a novel method of time series updation, and two dimensionality reduction approaches. Extensive experimental results are also presented demonstrating the effectiveness of all propositions.

## 7. Conclusion

Time series analysis of the index of several sectors of the economy of a country and the world can reveal several interesting points about the relationships of those sectors. A proper understanding of these relationships can prove critical in formulating several proactive economic policies in the current dynamic and extremely volatile economic conditions in the globe.

As an illustration of the effectiveness of time series analysis in investigating the behavior and association among various sectorial index, in this paper, we proposed a time series decomposition-based approach for deeper understanding and analysis of two sectors of the Indian economy – the Indian IT sector and the Indian CG sector. We have hypothesized that the Indian IT sector, which is strongly coupled with the world economic story, should reveal a strong association with the DJIA index and the exchange rate of the US Dollar to the Indian Rupees, while the Indian CG sector, being essentially coupled with the Indian economic story, should expose a strong association with the NIFTY index. We also contended that although the Indian IT sector essentially reflects the world economic story, the blue chips stocks of the IT sector have strong impacts on the NIFTY index and hence the Indian IT sector should also have a strong association with the NIFTY index. However, the Indian CG sector being essentially India-centric only, we expect a poor association between the Indian CG sector index with the DJIA index and the US Dollar to Indian Rupee exchange rate. Using our proposed decomposition approach and several statistical tests on correlation and cross-correlation among different time series we have validated all our hypotheses. After carrying out detailed analyses and association tests on the time series of various sectors, and a comprehensive validation of our all hypotheses, we have built a number of linear models among time series which were found to have strong association between them. In other words, we have constructed linear regression models between the Indian IT sector index with the DJIA index, the exchange rate of the US Dollar to the Indian Rupee and the NIFTY index. For the Indian CG sector, we have built a linear model with the NIFTY index. All the linear models of regression are built using suitably chosen training data set and their robustness and efficacies are tested using appropriate test data set. Forecast errors are computed for each month of the period of forecast so that the effectiveness of each model can be numerically evaluated.

The approach and methodology proposed in this paper are absolutely generic and can be applied to any sector of the world economy. Moreover, the results obtained using this approach can also be extremely useful for portfolio construction of stocks. By performing analysis on time series of several sectors and studying their seasonality characteristics, portfolio managers and individual investors can very effectively take decisions about buy/sell of stocks and appropriate timing. For speculative gains, the sectors exhibiting presence of dominant random components in their time series may be targeted.